\documentclass[twocolumn]{aastex63}
\usepackage[cmex10]{amsmath}
\usepackage{amssymb}    

\newcommand{\asec}{$^{\prime\prime}$}
\newcommand{\amin}{$^{\prime}$}
\received{11 Feb. 2025}
\revised{12 May 2025}
\revised{29 Jun. 2025}
\accepted{2 Jul. 2025}
\submitjournal{ApJ}
\shorttitle{WAT Radio Galaxy J1712$-$2435}
\shortauthors{Cotton et al.}
\newcommand*{\chg}{\textcolor{black}}
\newcommand*{\chgb}{\textcolor{black}}

\begin{document}

\title{MeerKAT 1.3 GHz Observations of the Wide Angle Tail Radio Galaxy J1712$-$2435}

\correspondingauthor{William Cotton}
\email{bcotton@nrao.edu}
\author[0000-0001-7363-6489]{W.~D.~Cotton}
\affiliation{National Radio Astronomy Observatory \\
520 Edgemont Road \\
Charlottesville, VA 22903, USA}
\affiliation{South African Radio Astronomy Observatory \\
Liesbeek House, River Park, Gloucester Road \\
Cape Town, 7700, South Africa} 

\author[0000-0001-9133-1005]{Gourab Giri}
\affiliation{Department of Physics, University of Pretoria,\\
 Private \chgb{B}ag X20, Hatfield, South Africa\\
}
\affiliation{INAF-Istituto de Radioastronomia, via P. Gobetti 101, I-40129, Bologna, Italy}

\author[0000-0000-0000-0000]{P.~J.~Agnihotri }
\affiliation{National Radio Astronomy Observatory \\
520 Edgemont Road \\
Charlottesville, VA 22903, USA}
\affiliation{San Francisco State University \\
San Francisco CA, USA} 

\author[0000-0002-4464-8023]{D.~J.~Saikia}
\affiliation{Inter-University for Centre for Astronomy and Astrophysics,\\
 Post Bag 4, Pune, 411007, India\\
}
\author[0000-0000-0000-0000]{K.~Thorat}
\affiliation{Department of Physics, University of Pretoria,\\
 Private bag X20, Hatfield, South Africa\\
}
\author[0000-0002-1873-3718]{F.~Camilo}
\affiliation{South African Radio Astronomy Observatory \\
Liesbeek House, River Park, Gloucester Road \\
Cape Town, 7700, South Africa}



\begin{abstract}
We present full polarization MeerKAT images of the wide-angle tail,
giant radio galaxy J1712$-$2435 at 1.3 GHz with
7.\asec5 resolution and an RMS sensitivity of 8 $\mu$Jy beam$^{-1}$.  
Due to the angular proximity to the Galactic Center
(l=359.6$^\circ$, b=+8.5$^\circ$) the immediate environment is not
well understood but there are massive clusters nearby.
Emission can be traced over an extent of 34.\amin6 which at the
redshift of 0.024330 corresponds to a projected length of 1.02 Mpc.  
The inner jets are quite straight but then bend and completely
decollimate into extended plumes nearly orthogonal to the initial jet
directions at a projected distance of approximately 100 kpc.  
The nearly unity brightness ratio of the inner jets suggest that they
are orientated within a few degrees of the plane of the sky. 
The 1400 MHz power is 3.9$\times 10^{24}$ W Hz$^{-1}$, somewhat below
the FRI/FRII divide.  
The total power emitted is estimated to be 5.6$\times 10^{41}$ erg
sec$^{-1}$ over the range 10 MHz to 100 GHz. 
The source dynamics are modeled with magneto-hydrodynamics
simulations; the result is a rough reproduction of the source's radio
morphology / appearance. 
This study further highlights the merit of alternative scenarios, 
calling for future observational and numerical efforts. 

\end{abstract}



\section{Introduction} 
The existence of radio galaxies with bent jets has long been known;
\cite{Owen_Rudnick_1976} describe a number of \chg{wide-angle tail} (WAT)
radio galaxies in observations of rich Abell clusters.
Subsequent VLA observations of a sample of WATs by \cite{ODonoghue1990},
with an analysis in \cite{ODonoghue1993} found them to occur mainly in
the centers of clusters without cool cores.
Bends and decollimation of the jets usually occur at similar
locations, \chg{tens} of kpc from the \chg{host} galaxy center.
The radio spectra steepen away from the galaxy core and down the jets
and tails.
The jet disruption and bends in these galaxies which are moving
slowly in their cluster environments were difficult to explain,
necessitating further insights from numerical simulations
\citep[e.g.,][]{Nolting2019,Nolting2022}. 

In a recent review of WAT radio galaxies\chg{,} \cite{ODea2023} describe 3
basic properties, 1) well collimated inner jets suddenly
decollimating into plumes or tails, 2) a bend in both jets to the
same side into either ``C'' or ``V'' shapes and 3) the plumes or tails not
being parallel. 
Most, but not all, WATs are in clusters or groups and tend to be
associated with the \chg{b}rightest \chg{c}luster \chg{g}alaxy (BCG).
Most \chg{known WATs} have radio \chg{powers} in a fairly narrow range of 10$^{42}$ to
10$^{43}$ ergs sec$^{-1}$ integrated over 10 MHz to 100 GHz.
This puts them near the FRI/FRII power divide \citep{Owen1994}.
\cite{ODea2023} speculate that the limited range of luminosities
results from a minimum needed for the jets to propagate the
\chg{tens} of kpc through the host and a maximum that allowed the jets to be
totally disrupted.
These authors also suggest that the decollimation occurs at the
transition from the host ISM to the cluster medium.

The nearby large WAT radio galaxy J1712$-$2435 was discovered during a survey
of the Galactic bulge described in \cite{Bulge} and was the subject of
followup observations and is discussed in the following.
Due to the proximity to the Galactic Center (l=359.6$^\circ$, b=+8.5$^\circ$), the extragalactic sky in this area has not been well studied.
The angular size of J1712$-$2435 (core component J1712427$-$243550) was
measured by \cite{Bulge} following around bends as 34.\amin6.
At a redshift of 0.024330 \citep{Allison2014} the corresponding
projected \chg{length} is 1.02 Mpc making this a \chg{g}iant
\chg{r}adio \chg{g}alaxy (GRG).
See \cite{LoTTS_GRG_2023} for a sample of GRGs up to 5 Mpc in
linear extent, and \cite{Dabhade2023JApA...44...13D} for a recent review. The largest known GRG extends to about 7 Mpc \citep{Oei2024Natur.633..537O}.

The observations \chg{and data reduction} are described in Section
\ref{Observations},  
the results \chg{are}  presented in Section \ref{Results}.
Dynamic modeling of the source morphology is described in Section
\ref{Modeling} and Section \ref{Summary} is a summary of the results.
\chg{The data products provided are listed in Section \ref{Products} .}

\section{Observations and \chg {Data Processing} \label{Observations}}
Observations were made of the WAT GRG J1712$-$2435 at L band (856 to
1712 MHz) on 2020-04-25 using the MeerKAT array in South Africa
\citep{Jonas2016, Camilo2018,DEEP2}.  
The observations were for 10 hours including calibration \chg{and
overhead} with 6 hours on source using 59 of the 64 antennas, 4096
spectral channels across the band, 8 second integrations and all four
of the products of the orthogonal linearly polarized feeds.
The pointing position was RA (\chg{J}2000) =
17\chg{h}12\chg{m}42.81\chg{s}, Dec(\chg{J}2000) = $-$24\chg{$^\circ$}35\chg{amin}
49.9\chg{asec} (l=359.592$^\circ$, b=8.518$^\circ$) and the project
code was SSV-20200423-FC-01.  
Processing followed the general approach of \cite{Bulge} and used the
Obit package\footnote{http://www.cv.nrao.edu/$\sim$bcotton/Obit.html}
\cite{OBIT}.

\subsection{Calibration}
Calibration and flagging of the data were as described in \cite{Bulge}
and \cite{MK_SMC}. 
The flux density/bandpass/delay/unpolarized calibrator was
PKS~B1934$-$638, 3C286 was the polarized calibrator and J1733$-$1304 was
used as the astrometric/gain calibrator. 
The flux density scale was set by the \cite{Reynolds94} spectrum of
PKS~B1934$-$638:
$$
  \log(S) = -30.7667 + 26.4908 \log\bigl(\nu\bigr)
  - 7.0977 \log\bigl(\nu\bigr)^2 $$
  $$+0.605334 \log\bigl(\nu\bigr)^3,$$
where $S$ is the flux density (Jy) and $\nu$ is the frequency (MHz).

\subsection{Imaging}
Imaging of J1712$-$2435 was performed in Stokes I, Q, U and
V as described in \cite{Bulge} and used Obit task MFImage
\citep{Cotton2018}.  
Imaging of Stokes I and V used 14 $\times$ 5\% fractional bandwidth sub-bands while
Q and U used 2\% sub-bands to allow detection of higher rotation
measure features; this gives 34 sub-bands across the band, many
blanked by RFI.
The field of view was fully imaged to a radius of 0.75$^\circ$ with
\chg{ facets on outlying strong sources} allowed to 1.5$^\circ$.
The Stokes I image was CLEANed to 30 $\mu$Jy beam$^{-1}$ and Q, U and
V to 20 $\mu$Jy beam$^{-1}$. A total flux density of
8.68 Jy was CLEANed in Stokes I.
The restoring beam used was 7.5\asec\ $\times$ 7.4\asec\ at position angle
$-$1.2$^\circ$. 
Peeling was used to reduce the effects of other bright emission in the
field; see Section \ref{peel}.
The off--source RMS in the broadband Stokes I image was 7.8 $\mu$Jy
beam$^{-1}$, 1.3 $\mu$Jy beam$^{-1}$ in Q and U and 4.0 $\mu$Jy
beam$^{-1}$ in V.

In order to approximately equalize the short baseline coverage, the
inner portion of the uv coverage had a Gaussian taper with a
\chg{standard deviation} of 500 $\lambda$ \citep{XGalaxy} applied.
This allows a more equal recovery of extended \chg{emission} across
the bandpass which results in better estimates of spectral index at
the cost of reducing sensitivity to the largest scale structure.

\subsection{Peeling\label{peel}}
The field of J1712$-$2435 contains a source \chg{(J171448-251435)} out
in the beam bright enough that residual artifacts from the
direction\chg{--}independent self\chg{--}calibration significantly
reduce the dynamic range. 
``Peeling'' \citep{Noordam2004} was used to reduce these artifacts. 
The emission from sources other than that to be peeled were
subtracted from the visibility data and the peel source was then self
calibrated. 
The response of the peel source was subtracted from the visibility
data using the gain solutions derived for it.

\subsection{Effective Frequency and Spectral Index}
The effective frequency of a broadband image depends on the details
of how the sub-band images were combined.
The bulk of the pixels in the resultant images 
\chg{have no reliably detectable signal in the subband images}
to fit for a spectrum, so the combination of the
sub-band images followed that in \cite{SMGPS,MK_SMC,Bulge}.
The combination of sub-bands used a weighted average where the
weighting of each sub-band was proportional to the average over all
mosaic images of 1/$\sigma$ where $\sigma$ is the off--source standard
deviation of the pixel values.
1/$\sigma$ is used rather than the more conventional 1/$\sigma^2$
because the steep spectrum of the Galactic \chg{foreground} causes the
latter weighting to essentially remove the effects of the lower
frequency sub-bands.
The same weights as used in \cite{Bulge} resulted in an effective
frequency of 1333.1 MHz.
Spectral indices were fitted by a nonlinear least squares fit in
pixels with
a broadband Stokes I brightness in excess of 200 $\mu$Jy/beam. 
A second version of the broadband Stokes I and spectral index image
which includes the error estimates from the least squares fitting is
also provided. 

\subsection{Faraday Analysis\label{Faraday}}
The Q and U cubes were subjected to a search in Faraday depth in each
pixel to determine the peak polarized intensity, rotation measure and
intrinsic polarization angle.
The search range in RM was $\pm$2500 rad m$^{-2}$ with an increment of
0.5 rad m$^{-2}$. 
A cube containing the derived parameters is produced.

\section{Results\label{Results}}
\subsection{Host Galaxy}
The core of J1712$-$2435 is located at RA(\chg{J}2000) = 17 12 42.85,
Dec(\chg{J}2000) = $-$24 35 50.2 which NED \chg{(NASA Extragalactic
Database\footnote{https://ned.ipac.caltech.edu/})}
associates with WISEA
J171242.78$-$243547.9, 2MASX J17124278$-$2435477.    
The redshift of the host galaxy from HI measurements is 0.024330 $\pm$
3.00e-6 \citep{Allison2014}, or velocity=7294 km/sec. 
\cite{Hasegawa2000} also gives a velocity of 7294 km/s and classifies the
host as a bright elliptical 0.\amin25 in angular size.
At the distance of this galaxy 1\asec\ corresponds to 0.493 kpc
\citep{Cosmology_Calculator_2006}\footnote{The cosmology assumed was
  H$_0$=69.6, $\Omega_M$=0.286 and $\Omega_{vac}$=0.714}. 

\subsection{Morphology}
J1712$-$2435 has well collimated inner jets, with bends, emerging from the
core. 
Both jets become decollimated and develop extended plumes mostly
orthogonal to the original jet directions.
The projected northern jet goes a shorter distance before
decollimating than the southern jet which has several bends before
becoming decollimated. 
The northern plume breaks up into numerous filaments extending to the
east; the southern plume has northern and southern extensions.
\chg{The longest of these filaments is about 700\asec\  or a projected
length of $\sim$345 kpc.}

\begin{figure*}
\centerline{\includegraphics[width=6.8in]{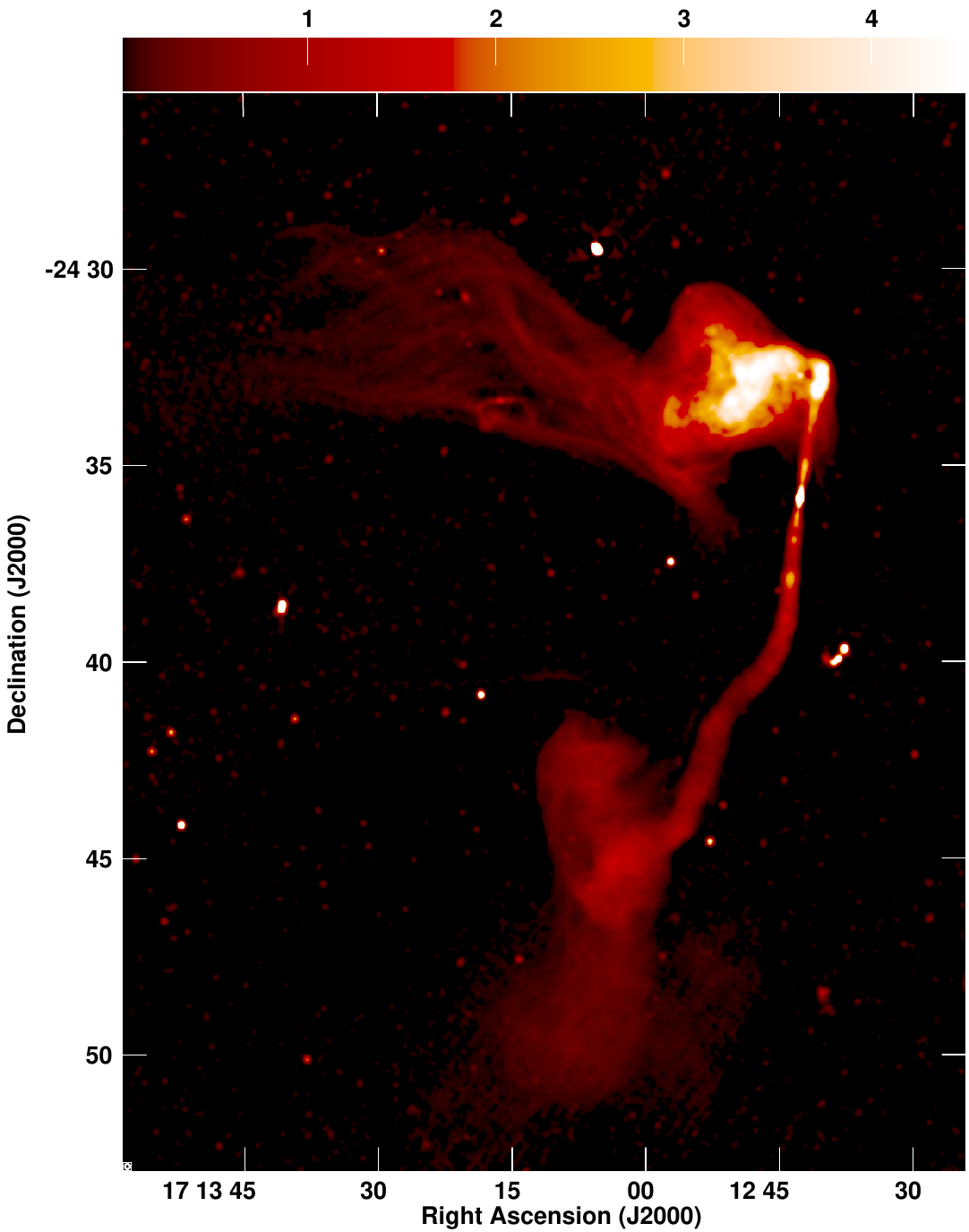}}
\caption{\chg{Stokes I image} of J1712$-$2435; the \chg{spectral} flux
  density scale in mJy beam$^{-1}$ is given in the scale bar at the top.
\chg{A representation of the FWHM of the psf} is shown in the box in
the lower left corner. 
}
\label{fig:Full_IPol}
\end{figure*}

The northern jet bends and decollimates at a projected distance of
98 kpc from the nucleus.
The first bend in the southern jet is also at a projected distance of
97 kpc but it only decollimates at a distance of 277 kpc from the
nucleus. 
Away from the core, both jets become well resolved in the transverse
direction.
The Stokes I structure is illustrated in Figure \ref{fig:Full_IPol}.

\chg{The ``C'' shaped structure of the source shows that there is
some force that causes the jets to bend in the same direction and then
disrupt.
The straight inner jets show that the bending effect increases with
distance from the galactic nucleus.
An external wind is one such force.
The denser ISM in the inner regions of the host would reduce the
effects of the wind allowing the jets to be straight until the jet
reaches the outer part of the galaxy.
This is consistent with both jets in this source remaining straight to
a projected distance of 100 kpc.
Such a scenario has long been suspected.
The principle counter--argument has been that such radio sources are
frequently in the Brightest Cluster Galaxy (BCG) of the cluster.
As such, they define the bottom of the cluster potential well and, in
a relaxed system, should be at rest with the intracluster medium
\citep{ODea2023}. 
However, in the case of merging clusters, members of one cluster
``falling into'' the potential of the other could experience a
significant wind.
}

\subsection{Spectral Properties}
\begin{figure*}
\centerline{\includegraphics[width=6.8in]{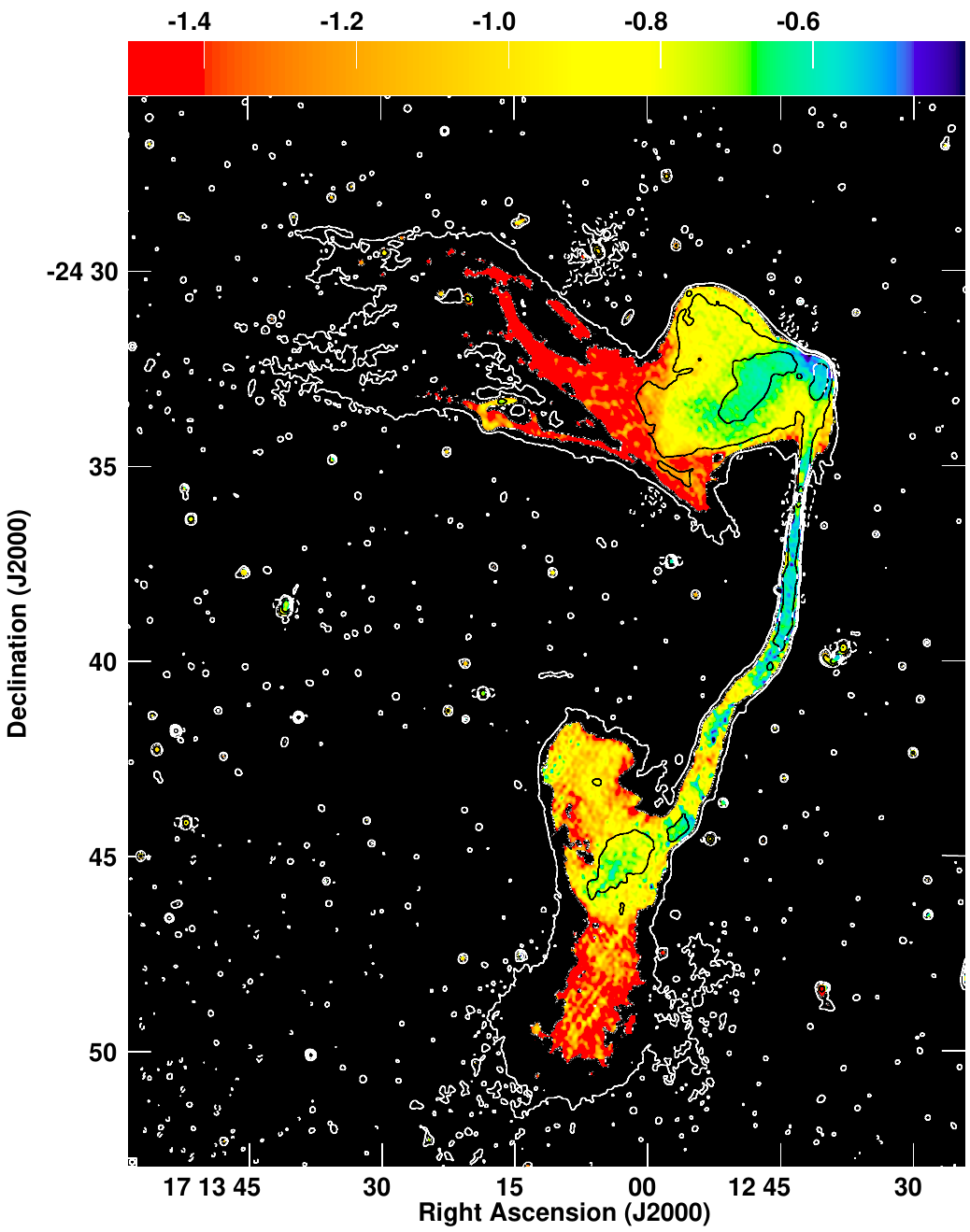}}
\caption{The spectral index of J1712$-$2435 as displayed in Figure
  \ref{fig:Full_IPol}, with values given in the bar at the top.
Stokes I contours are shown at levels of $\pm$ 50 $\mu$Jy beam$^{-1}$
and powers of 4 above that.
}
\label{fig:Full_SI}
\end{figure*}

The integrated broadband flux density is \chg{2.764 $\pm$ 0.0014} Jy
at 1333.13 MHz or a \chg{radio} luminosity at 1400 MHz of 
\chg{3.924 $\pm$ 0.002}$\times 10^{24}$ W Hz$^{-1}$,
somewhat below the FRI/II divide \citep{Owen1994} of $\sim 10^{25}$ 
\chg{W Hz$^{-1}$} at 1.4 GHz. 
Assuming a spectral index of $\alpha$=$-$0.7, the power emitted between
10 MHz and 100 GHz 
is 5.6$\times 10^{41}$ ergs sec$^{-1}$.
The spectral index distribution is shown in Figure \ref{fig:Full_SI}.
The spectrum is flat on the core and steepens along the jet becoming
very steep in the outer regions of the plumes.

\subsection{Polarization and Magnetic Field}
\begin{figure*}
\centerline{\includegraphics[width=6.8in]{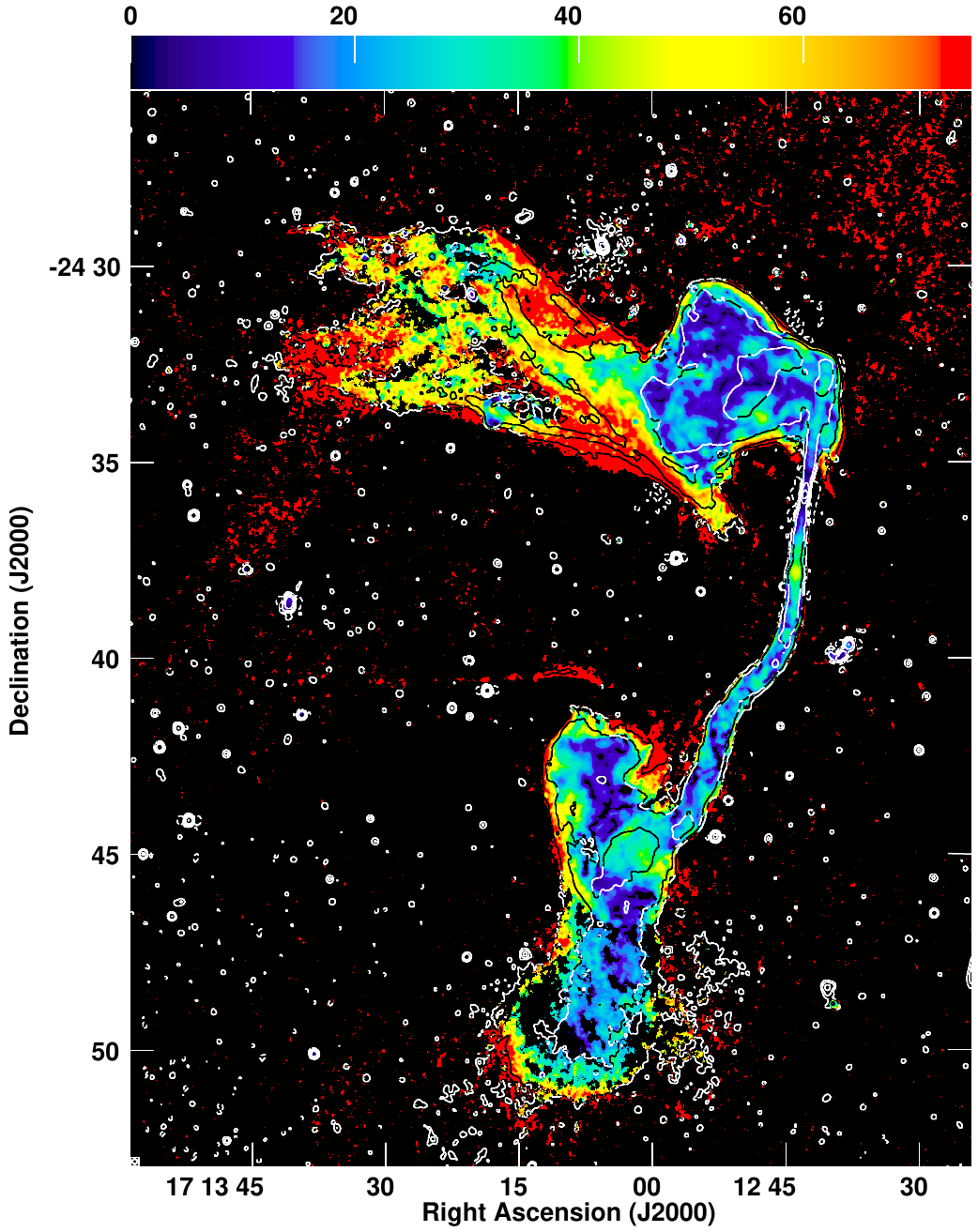}}
\caption{The fractional linear polarization of J1712$-$2435 as displayed in Figure
  \ref{fig:Full_IPol}.
The fractional polarization in percent is shown in color with values
given in the bar at the top.
Stokes I contours are shown at levels of $\pm$ 50 $\mu$Jy beam$^{-1}$
and powers of 4 above that.
}
\label{fig:Full_FPol}
\end{figure*}

\begin{figure*}
\centerline{\includegraphics[width=6.8in]{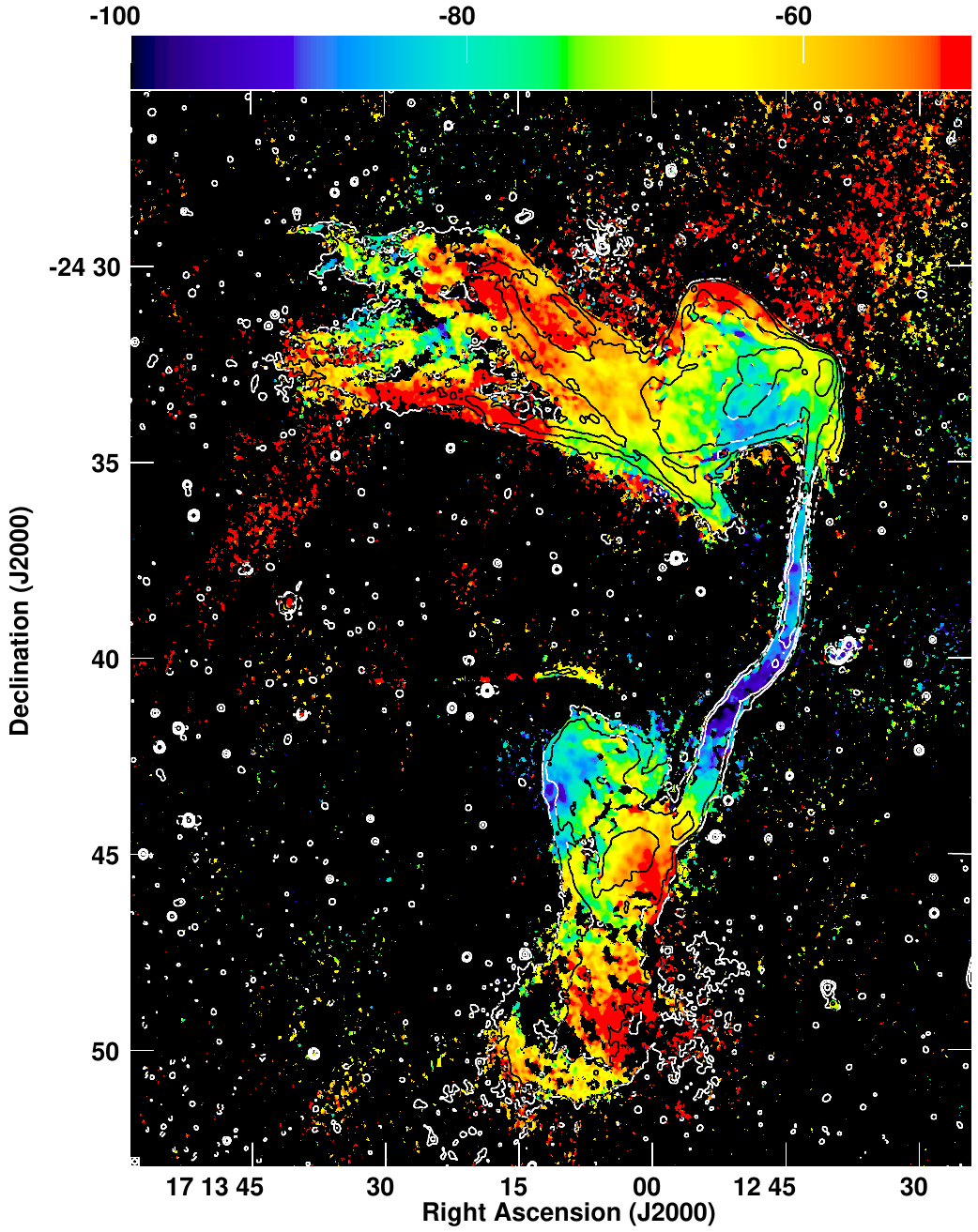}}
\caption{The peak Faraday depth of J1712$-$2435 as displayed in Figure
  \ref{fig:Full_IPol}.
The rotation measure in rad m$^{-2}$ is shown in color with values
given in the bar at the top.
Stokes I contours are shown at levels of $\pm$ 50 $\mu$Jy beam$^{-1}$
and powers of 4 above that.
}
\label{fig:Full_RM}
\end{figure*}

The fractional polarization is shown in Figure \ref{fig:Full_FPol}.
The rotation measure determined from the peak unwrapped polarized
intensity is shown in Figure \ref{fig:Full_RM}.
The variations of Faraday rotation are relatively smooth and do not
cover a large range.
\chg{The medium surounding the source does not contain a dense, clumpy
magnetized plasma as is seen in some clusters \citep{Rudnick2023}.}
The inner jet has a larger negative value in the longer southern  jet
than in the northern jet.
Much higher values could have been detected.


\begin{figure}
\centerline{\includegraphics[width=2.0in]{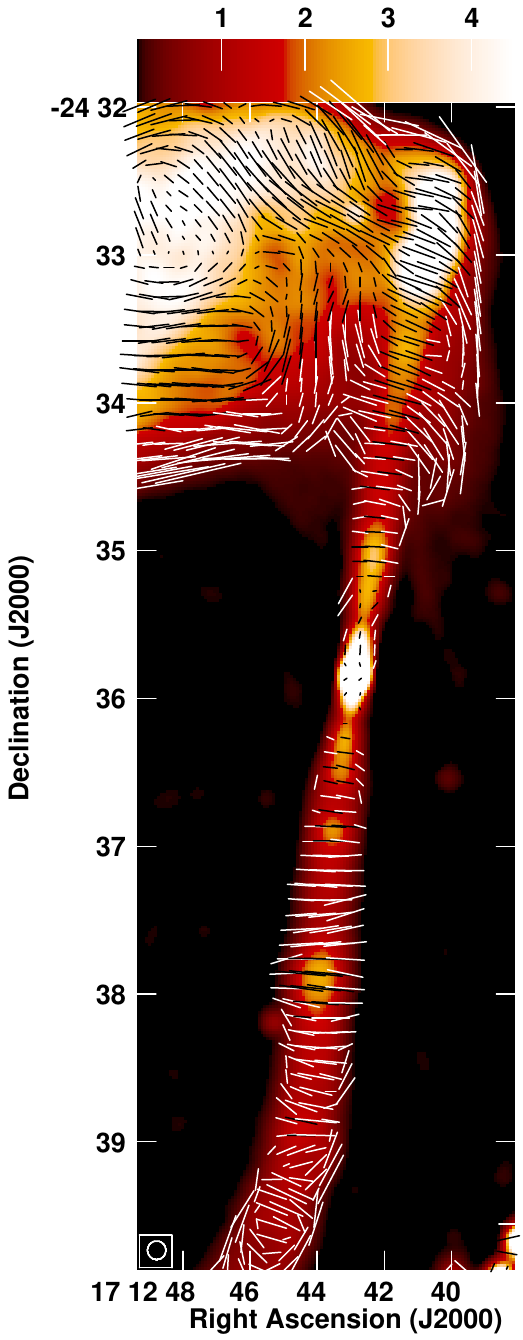}}
\caption{The inner jet of J1712$-$2435 with Stokes I shown in color with
  the scale bar at the top in mJy beam$^{-1}$ and fractional
  polarization ``B vectors'' showing the orientation of the
  magnetic field.
}
\label{fig:Inner_Jet_pol}
\end{figure}

\begin{figure}
\centerline{\includegraphics[width=3.5in]{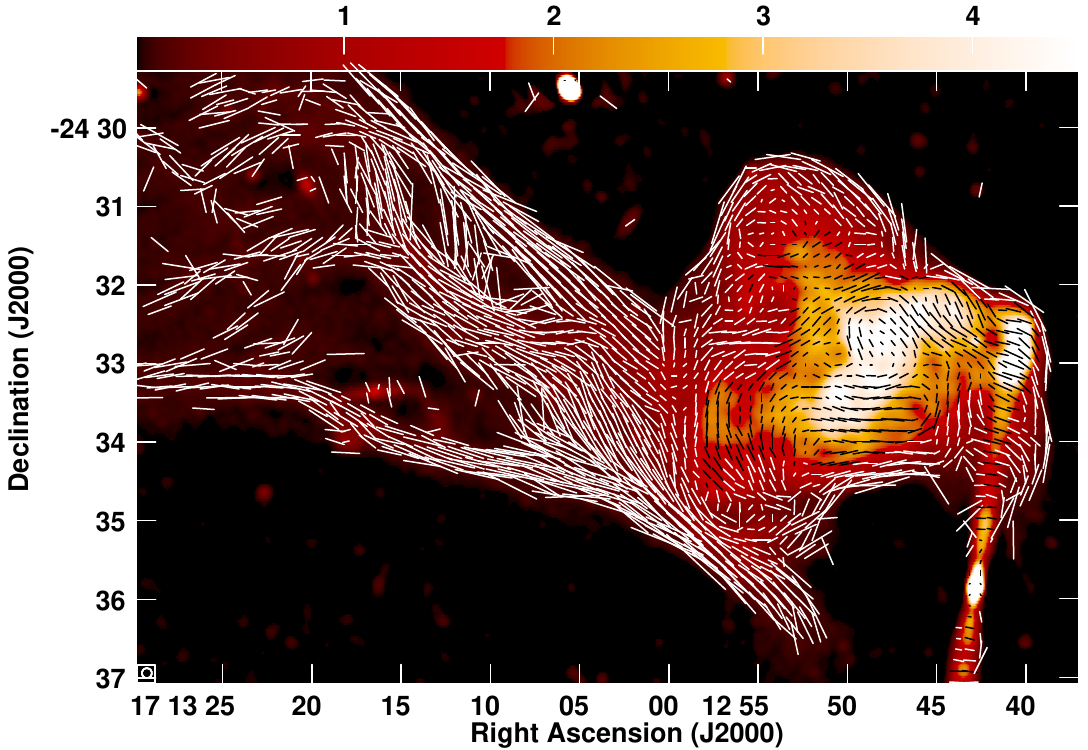}}
\caption{The northern plume of J1712$-$2435 with Stokes I shown in color with
  the scale bar at the top in mJy beam$^{-1}$ and fractional
  polarization ``B vectors'' showing the orientation of the
  magnetic field.
}
\label{fig:North_pol}
\end{figure}

\begin{figure}
\centerline{\includegraphics[width=3.5in]{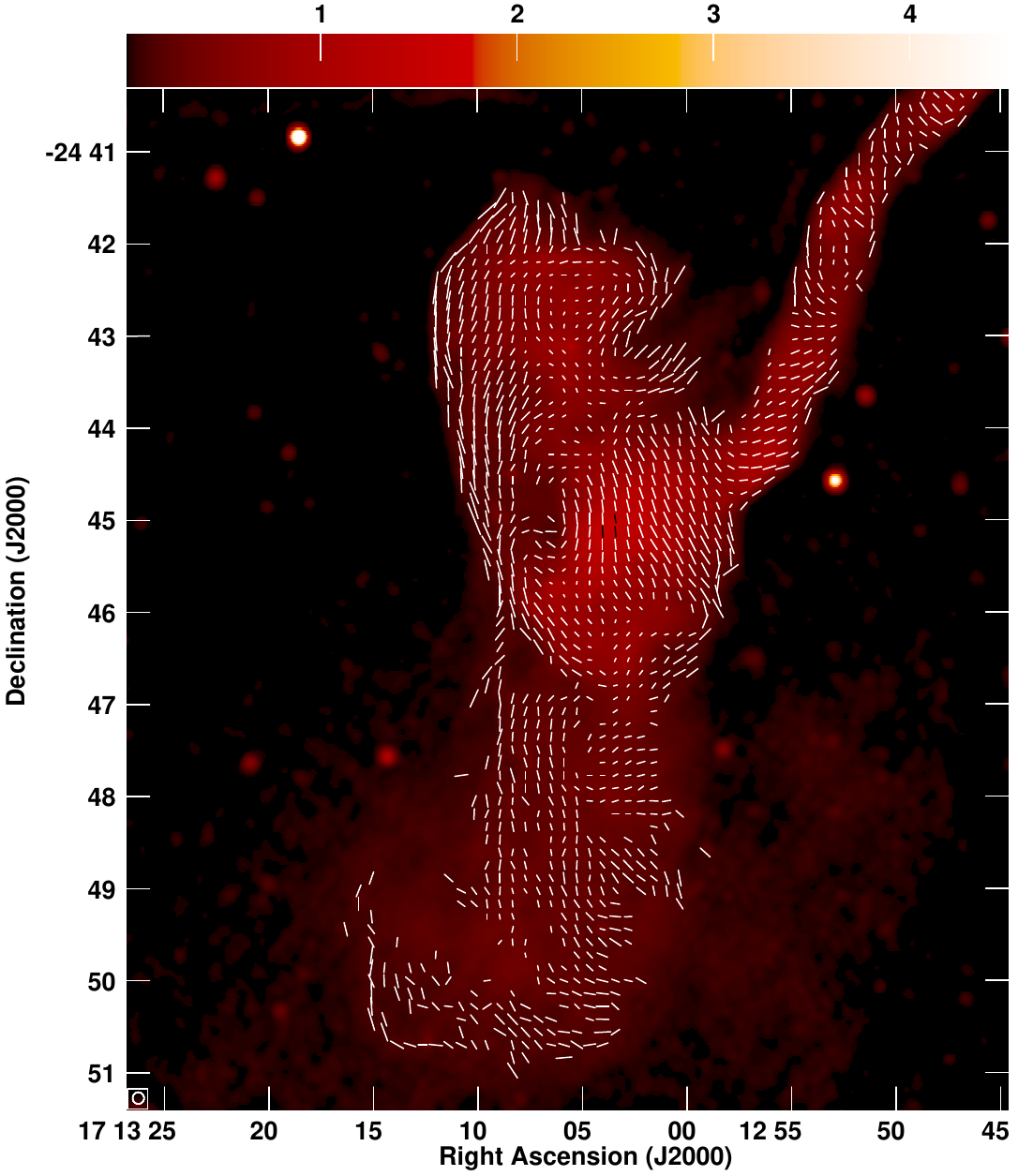}}
\caption{The southern plume of J1712$-$2435 with Stokes I shown in color with
  the scale bar at the top in mJy beam$^{-1}$ and fractional
  polarization ``B vectors'' showing the orientation of the
  magnetic field.
}
\label{fig:South_pol}
\end{figure}

The magnetic field direction in the inner jet
(Figure \ref{fig:Inner_Jet_pol}) is transverse to the jet flow. 
The magnetic field in the southern jet becomes dominantly longitudinal
before the jet decollimates. 
The field orientation in the jets is consistent with what is seen in
FRI sources, where the field is either predominantly perpendicular to
the jet axes or has a combination of perpendicular and parallel
components \citep[for reviews,
see][]{Bridle1984ARA&A..22..319B,Saikia2022JApA...43...97S}. 
The magnetic field structure in the northern plume is shown in Figure
\ref{fig:North_pol}. 
The magnetic field wraps around the outer edge of the bright region
but then largely follows the filaments away from the main axis of the
source, \chg{showing these to be magnetic structures.}
Note, there is a head/tail radio galaxy superposed on the southernmost
filament with its nucleus at RA=17 13 16.41, Dec=$-$24 33 21.2.
The magnetic field structure in the southern plume (Figure
\ref{fig:South_pol}) is less organized except where the jet enters the
plume and the field is transverse to the direction of the jet  and is
largely along the direction of the northern extension of the plume.
The field wraps around the edges of the plume.
The rim of very high fractional polarization around the southern lobe
seen in Figure \ref{fig:Full_FPol} is very likely an artifact. 

\subsection{Host Galaxy and Environment}\label{Sec:Host Galaxy and Environment}
The morphology of the source strongly suggests that it is being
subjected to gas flows in an intergalactic medium as is frequently
observed in clusters of galaxies.
J1712$-$2435 appears quite close to the Galactic Center,
l=359.6$^\circ$, b=+8.5$^\circ$, so its immediate environment is not
well understood. 
This galaxy is \chg{2 1/2 degrees from} the direction of the Ophiuchus super
cluster \citep{Hasegawa2000} and near, but not in, the Sagittarius cluster
\citep{Djorgovski1990} at a velocity of $\sim$8600 km s$^{-1}$\chgb{.}

\cite{FoF} associate the hosts of J1712$-$2435 and the nearby bright
FRII AGN J1713$-$2502 with a group of 10 galaxies, their group number
6489, centered at RA=17 12 31.75, Dec=$-$24 45 05.4 with a redshift,
z=0.02847 (\chg{v} = 8535 km s$^{-1}$) and the $\sigma$ of the velocity
dispersion of 751 km s$^{-1}$.
J1713$-$2502 is the brightest source in the field of J1712$-$2435 and its
core is located at RA=17 13 50.50, Dec=$-$25 02 27.5 which NED associates
with WISEA J171315.44$-$250226.8, 2MASX J17131541$-$2502266.
\cite{Allison2014} give the redshift of J1713$-$2502 as 0.0285 $\pm$
1.67e-4 (vel=8566 $\pm$ 50 km s$^{-1}$).
J1713$-$2502 is smaller in physical size than J1712$-$2435 ($\sim$
5\amin\ \chg{$\approx$} 150 kpc end-to-end)  and shows no signs of bends. 
Six of the ten members of Group 6489 are detected in the MeerKAT
image (visible in FITS file J1712-2435\_I\_FitSpec.fits.gz, see Section \ref{Products}).
J1712$-$2435 is \chg{apparently} located in an over dense region of galaxies.


\subsection{Jet/Counterjet Brightness Ratio}
If the jet has relativistic bulk motion, the Doppler boosting of the
jet and deboosting of the counterjet will result in a ratio \chg{$R$} of the
jet/counterjet brightness ratio \citep{Giovannini2004}:
$$R\ =\ (1+\beta cos\theta)^{(2-\alpha)} (1-\beta
cos\theta)^{-(2-\alpha)},$$
where $\beta$ \chg{is the ratio between the jet bulk flow speed and
  the speed of light},
$\theta$ is the angle between the line of sight to the source and the
jet direction and $\alpha$ is the spectral index.
There are detailed fluctuations on both the inner jet and counterjet
so the brightness ratio was estimated from the ratio of the average
brightness in 75 kpc long slices down the midline of the jet and
counterjet. 
The derived brightness ratio (north/south) is 1.14 \chg{$\pm$ 0.045}.
To produce a ratio so close to unity, the jets must be either very
slow, or, quite close to the plane of the sky.
Given the large extent of the radio source, a slow jet seems
implausible. 
In their discussion of the velocities of jets measured in GRGs,
\cite{ODea2023} give a range of 0.2 to 0.7 c.
A jet moving at 0.2 c would have to be within 7$^\circ$ of the plane
of the sky to have a jet/counterjet ratio as close to unity as 
measured.
We conclude that the jet is within a few degrees of the plane of the
sky and cannot meaningfully constrain the velocity.

Although the oppositely-directed jets are quite symmetric in
brightness, a remarkable feature of these jets is the asymmetry in
their lengths. For a small sample of WATs \cite{ODea2023} find the
median value of the ratio of oppositely-directed jet lengths to be
1.2, with the highest value being 1.8, and the ratio being similar
with increasing jet length. For J1712$-$2435 the ratio of the jet
lengths is much larger with a value of about 2.8. Since the
oppositely-directed jets in J1712$-$2435 are quite collinear 
\chg{(within 3$^\circ$)}, it is
reasonable to assume that they are inclined at similar angles to the
line of sight. 
\chg{The large length ratio may be the result of} an asymmetry in the
external environment, 
assuming the jets to be intrinsically symmetric. 
As discussed in Section \ref{Modeling} on
numerical modeling of the source dynamics, the large jet length ratio
\chg{may be} due to a highly variable intergalactic environment, which would be
useful to probe from deep X-ray observations.  

\subsection{\chg{Jet Expansion}}
\begin{figure}
\centerline{\includegraphics[width=3.5in]{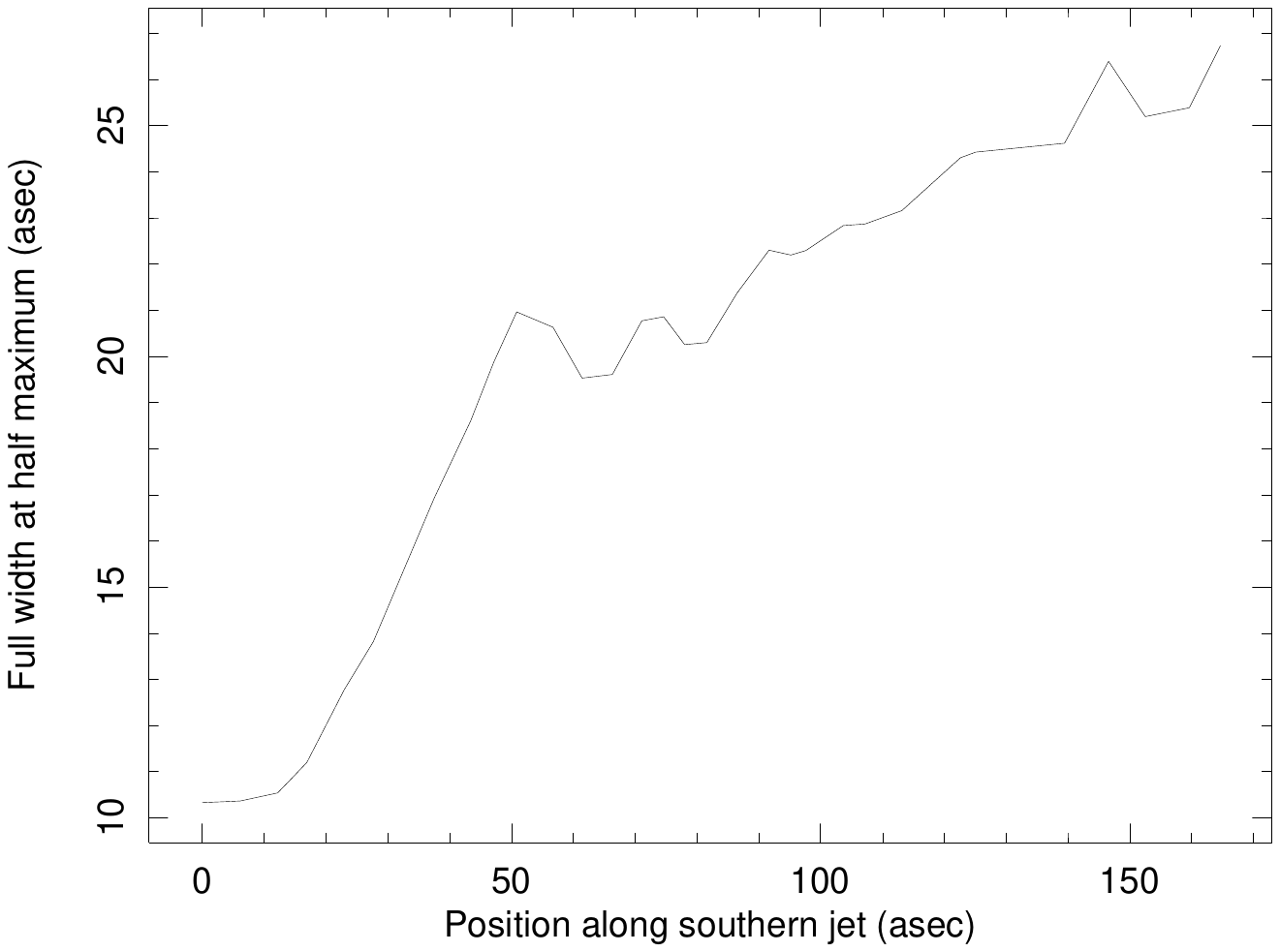}}
\caption{\chg{The width of the straight portion of the southern jet as FWHM
  as a function of distance from the nucleus.}
}
\label{fig:JetExpansion}
\end{figure}

\chg{One of the diagnostics of a jet's velocity and surounding medium is
how the jet expands with distance from the nucleus.
Much of the northern jet overlays the northern lobe so difficult
to separate jet from lobe.
However the inner portion of the southern jet is quite isolated and
Gaussian widths were fitted at various locations.
This is shown in Figure \ref{fig:JetExpansion}.
After an initial section where the jet was unresolved, it expands
rapidly until about 50\asec (25 kpc) at which point it abruptly slows
to a lower rate.}



\section{Numerical Modeling of Source Dynamics \label{Modeling}}
Given the complexity of J1712$-$2435, we aimed to develop a numerical model for this source, building upon inferences drawn from the observational constraints discussed above. Several key points have been taken into special consideration for developing the model, as follows. The tailed morphology of J1712$-$2435 suggests an intricate association with gas flow in the surrounding intergalactic medium (IGM), given its location in an overdense, galaxy group-like environment. \chg{The bending topology of the two jet arms differs both morphologically and in extent, suggesting that asymmetric IGM wind behavior is likely responsible for their formation. Given the complex jet-environment interactions involved under such assumption, we proceeded to model each arm separately, allowing for model flexibility.} The two arms are fairly well-collimated and straight up to nearly equal extents on either side, suggesting that the IGM wind is nearly ineffective in this initial phase of jet flow, likely due to the host galaxy's gas distribution partially protecting the jet flow \citep[see, e.g., the extent of optical-line emitting gas, and hot coronal gas in early-type galaxies;][]{Forman1985,Baum1989}. To formulate such a complex scenario in a numerical grid, we introduced a rotating IGM with a constant angular speed, resulting in a lower linear wind speed near the jet-injection and intial propagation zone, with increasing speed as the distance increases. \chg{We note that the angular rotation used here was a computational device to simulate the shielding effect of the host galaxy gas, not a model of the IGM. This approach replaced variable shielding (which is difficult to model numerically) with a variable velocity (easier to implement numerically).} Finally, we keep in mind that the source has been estimated to result from a \chg{low-power} jet flow evolving nearly in the plane of the sky.

\subsection{Simulation Setup}
We used the PLUTO astrophysical code for our modeling \citep{Mignone2007}, a tool now widely employed for simulating magnetohydrodynamical processes across various astrophysical systems. We applied standard module configurations for the simulations, including the HLLC Riemann solver and a second-order \chg{Runge–Kutta} solver for spatiotemporal evolution in a Cartesian system. Additionally, we implemented divergence cleaning for the magnetic field and selected the \chg{Taub--Matthews} equation of state to model relativistic flow \citep{Mignone2005,Mignone2006}.
\subsubsection{Ambient Environment}
We formulated a galaxy group medium using the King's density profile \citep{Cavaliere1976}, defined as follows,
\begin{equation}
    \rho (r) = \rho_{0} \left[1+\left(\frac{r}{r_c}\right)^2\right]^{-3\beta/2}
\end{equation}
where the density varies with radius $r = \sqrt{x^2 + y^2 + z^2}$. By setting the core radius ($r_c$) to approximately 33 kpc, the parameter $\beta$ to 0.55, and the core density ($\rho_0$) to 0.001 amu/cc ($\sim 180$ times \chg{today's} critical density), the total mass approaches a few times $10^{12} M_{\odot}$ \citep[e.g.,][]{Mulchaey2000}, with an equivalent virial radius \citep{Gaspari2020} of $\sim 400$ kpc \citep[relevant to e.g.,][]{Oei2023}. \chg{The above choice of parameters is also motivated by several recent studies that have numerically modeled a representative large-scale environment similar to ours \citep[e.g.,][]{Hardcastle2018,Musoke2020,Giri2025}.} The pressure profile $P(r)$ follows the density profile $\rho(r)$, resulting in an isothermal atmosphere at 1.2 keV \citep{Sun2009,Lovisari2015}. \chg{To establish an initial stable equilibrium for the ambient medium, we impose a gravitational acceleration profile ($\mathbf{g}$) derived from the condition of hydrostatic equilibrium, $\nabla P = \rho \mathbf{g}$, thereby ensuring that the plasma remains confined within the computational domain against pressure-driven expansion. However, once the jet-environment motion is introduced, the ensuing plasma dynamics are primarily governed by the jet–environment interaction, with gravity exerting only a secondary influence on the system's evolution.}

As previously noted, we introduced rotational motion to the ambient medium with an angular velocity $\omega$. For simplicity, rotation is applied only in the $x-$ and $y-$ directions of the Cartesian system, with the rotation axis aligned along the (fixed) $z-$axis. The resulting velocity profile is defined as follows,
\begin{equation}
    (v_x, \, v_y, \, v_z) \,  \equiv \, \omega \, R_{\rm rot} (- \, {\rm sin} \, \phi_{\rm rot}, {\rm cos} \, \phi_{\rm rot}, 0)
\end{equation}
Here, $R_{\rm rot}$ infers to the radius defined in the $x-y$ plane ($= \sqrt{x^2 + y^2}$), and $\phi_{\rm rot}$ refers to the angle in the $x-y$ plane ($= {\rm tan}^{-1} y/x$). The choice of $\omega$ is crucial here. Due to the asymmetric structure of the north and south jet arms, we adopted different $\omega$ values to model each arm separately. For the southern arm, where bending is less drastic, we used a lower $\omega$ value of $0.01c/250 \, \text{kpc}$ (equivalent to a linear speed of $0.01c$ at 250 kpc; $c$ being light speed). In contrast, the northern arm, which shows a sharp deflection in the lobe, required a higher $\omega$ value of $0.015c/250 \, \text{kpc}$ (equivalent to a linear speed of $0.015c$ at 250 kpc). The linear wind speeds highlighted here correspond to values observed in non-relaxed galaxy cluster atmospheres and are associated with turbulent eddies and bulk flows on scales of hundreds of kiloparsecs \citep{Burns1998}.

\chg{A recent review by \citet{ODea2023}, suggests that the formation of wide-angle tail sources typically requires wind speeds exceeding $1000$ km/s, implying a disturbed or merging galaxy group or cluster environment. Given that our source J1712--2435 exhibits more extreme bending features, characteristic of an even more dynamically influenced system, it is reasonable to infer that even higher wind speeds are necessary \citep{Burns2002}.  Supporting this inference, recent simulation work on the bent double radio jets of 3C 75 \citep{Musoke2020} demonstrates that stronger jet bending and lobe deflection are induced with substantial wind speeds on the order of 0.01c.}

We note that we have rigorously tested multiple variations of $\omega$ in separate simulation runs but found that these selected values best reproduced the observed structure of J1712$-$2435 (alternative values \chg{were less able to reproducing the source's morphology}, which we therefore choose not to showcase here).


\subsubsection{Jet Injection}
Given our velocity profile of the environment, which flows from the (negative) $x-$axis to the (negative) $y-$axis \chg{(counterclockwise)}, we chose to inject our one-sided jet flow along the negative $x-$direction for both the northern and southern jet-arm simulations. It is important to note that this orientation is simply a choice that we made; with our formulated spherical ambient medium, injecting the jet in a different direction would yield equivalent results without impacting the outcome.

We also note that we simulate a one-sided jet to account for the impact of the rotating ambient medium, allowing us to capture the sheltering effect of the host galaxy's gas on the initial jet flow. At the same time, modeling a two-sided jet under these conditions would likely result in the formation of an S-shaped radio galaxy rather than a tail structure. Furthermore, the observed asymmetry between the two arms suggests a highly variable intergalactic environment. Accurately modeling such a turbulent and dynamic intergalactic medium in a single simulation run poses significant computational and methodological challenges, making the current approach the most feasible and effective under these constraints.

We inject the jet through a small cylindrical region at the center of the environment, with a radius corresponding to the jet radius ($R_{\rm jet}$), chosen to be 500 pc, and a length of 1.5 kpc. The jet velocity is directed along the negative x-axis and is controlled by a \chg{bulk} Lorentz factor ($\Gamma$) of 5. The jet is underdense relative to the surrounding group environment by a factor of $10^{-5}$, \chg{giving $\rho_{\rm jet} = 10^{-5} \times \rho_0 $} \citep{Rossi2017,Giri2025}. The jet is nearly pressure-matched with the ambient medium. We inject the jet with a toroidal magnetic field \citep[a commonly used configuration;][]{Mignone2010} defined as,
\begin{equation}
    B_y = B_{\rm jet}\, \mathcal{R} \, {\rm sin}\, \theta, \quad B_z = -B_{\rm jet}\, \mathcal{R} \, {\rm cos}\, \theta
\end{equation}
where, ($\mathcal{R}, \theta$) is the polar coordinate in the perpendicular plane to the jet-flow direction, i.e., in the $y-z$ plane. The magnetic field strength is regulated by the parameter $\sigma$, which represents the ratio of Poynting flux to enthalpy flux, set to 0.01 \citep[typically assumed at the jet injection sites;][]{Rossi2017,Mukherjee2020}.

The selection of the above parameters was guided by the requirement that the injected jet power should be \chg{lower than the FR I/FR II power divide}, in order to meet the observational constraints. The kinetic energy of the injected jet has been estimated \citep{Rossi2017} to be $6.57 \times 10^{43} \, \text{erg/s}$, classifying the jet as a low-power FR-I type. This is also supported by the fact that the radio power of the evolved structure is typically \chg{a few times to a couple of} orders of magnitude lower than the mechanical power of the jet \citep{Birzan2004,Hardcastle2018}. We will observe the evolution of the structures along the $z-$axis, monitoring their progression in the plane of the sky. 

We introduced a passive tracer variable, labeled as `tr1' ($Q_k$), at the point of jet injection, which evolves according to the advection equation in its conservative form, as described below \citep{Matthews2019},
\begin{equation}
    \frac{\partial(\rho Q_k)}{\partial t} + \nabla \cdot (\rho Q_k v) = 0
\end{equation}
This tracer acts as a `color' effectively tied to the jet's density and velocity $(\rho, \, v)$, allowing us to track the flow of jet material as the simulation progresses. As the tracer evolves alongside the jet, its value in any grid cell within the simulation box represents the fraction of jet material present at that location, ranging from 0 (indicating no jet material) to 1 (indicating the cell is completely filled with jet material). This approach provides a powerful tool for visualizing and analyzing the distribution and mixing of jet material with the surrounding medium, offering deeper insights especially into the dynamical processes that govern jet propagation in an evolving ambient environment. The tracer values can help identify areas with high concentrations of jet-beam material (tracer values typically greater than $0.5$) as well as the jet-lobe regions with lesser concentration (tracer values lower than $0.5$) \citep[see, e.g.,][]{Rossi2008,Mukherjee2020}.

The computational domain for simulating the southern jet spans (in units of 50 kpc) $-8.1 \leq x \leq 0.1$, $-4.0 \leq y \leq 1.0$, and $-1.8 \leq z \leq 1.8$, distributed across a grid of $[820 \times 500 \times 360]$ cells. For the northern jet, the domain spans (in units of 50 kpc) $-6.1 \leq x \leq 0.1$, $-5.0 \leq y \leq 1.0$, and $-1.8 \leq z \leq 1.8$, with a grid distribution of $[620 \times 600 \times 360]$ cells. We designated the `southern' and `northern' jets based on the observed structure of J1712$-$2435; however, as mentioned previously, the one-sided jet flow in our simulations are directed along the negative $x-$axis.

\subsection{Progression of Bent-tail Features in Jet Morphology}
The underdense jet material, when subjected to the flow of the denser ambient medium, is prone to bending. Below, we elaborate on the adopted morphological features of the propagating jets, highlighting their distinctions resulting from slight changes in the environmental wind speed.

\subsubsection{Development of the Southern Jet Structure}
Following the onset of jet injection, the jet beam rapidly expands in a straight path due to its initial high propagation speed, which counteracts the influence of the lower linear velocity of the environmental gas flow. During this phase, backflow material is produced from the jet head due to the well-known pressure imbalance mechanism at the jet head-ambient medium interface \citep{Leahy1984,Bromberg2011}. This backflow eventually forms a thin cocoon as a result of the jet's rapid expansion \citep{Cielo2017}. Since the cocoon consists of lower kinetic energy and diluted material compared to the jet beam, it becomes susceptible to slight deflection along the direction of the environmental flow. 

This phase continues for approximately 29 Myr, during which the jet beam remains nearly linear, accompanied by the tilted narrow cocoon, until the stronger ambient flow begins to significantly affect the now-decelerated jet head, causing it to bend along the ambient medium's flow direction. This bending initiates around a distance of 200 kpc, where the environmental wind speed reaches $\sim 0.008c$. Over time, the distortion at the jet head intensifies, as does the evolving cocoon, with the entire jet material now channeling into one side, gradually developing a laterally expanding lobe-structure (perpendicular to jet-flow direction) beside the preexisting tilted cocoon (aligned with the jet-flow). This progressively develops into a bifurcated lobe-cocoon feature that continues to grow with time.

\begin{figure*}
\centerline{\includegraphics[width=2\columnwidth]{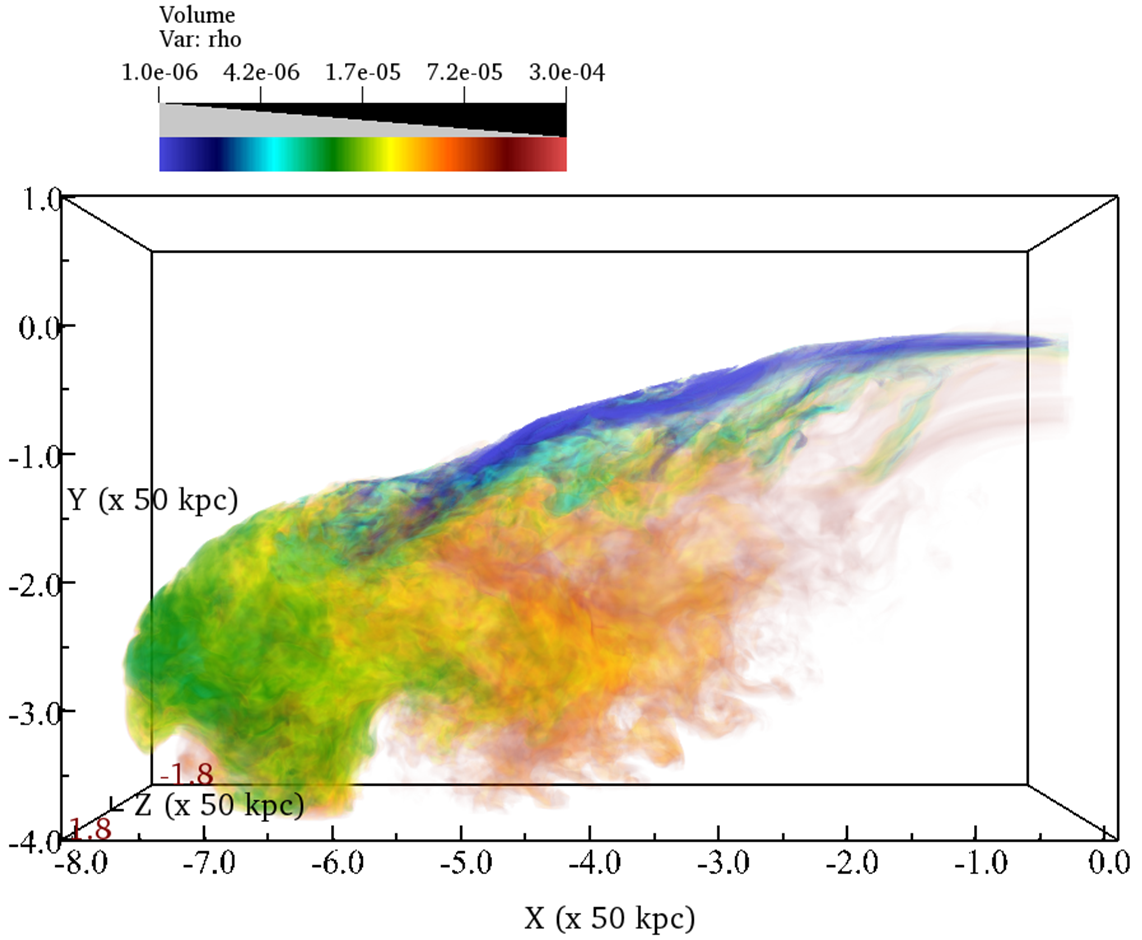}}
\caption{A simulated 3D volume-rendered density map illustrating the evolved jet-lobe structure at 75 Myr (dynamical age), \chg{resembling} the southern arm of J1712$-$2435. The jet beam (in blue) decollimates at approximately 275 kpc after bending twice due to the environmental wind flow ($\omega = 0.01c/250 \, \text{kpc}$). The lobe structure also develops bifurcated extensions. The density color scale is presented relative to 0.001 amu/cc, with an attached opacity scale to the colorbar (for better visualisation).}
\label{fig:south_arm}
\end{figure*}

Following the first bending, the jet head attempts to now realign its flow direction (around 52 Myr), producing a short, straight path before bending again at approximately 250 kpc (around 57 Myr), \chg{when the environmental wind speed reaches $0.01c$ \citep[see also,][]{Musoke2020}.} The spatial extent of this intermediate linear propagation phase (we attribute it as `sandwiched layer') varies significantly during the jet's evolution (typically spanning tens of kpc), further underscoring the varying nature of the emerging structure. 

Interestingly, this multiply curved morphology resists rapid decollimation and disruption due to the magnetic field tightly coiled around the jet beam, which stabilizes the flow. This behavior is reproduced in simulations, such as by \citet{Upreti2024} \citep[see also,][]{Mukherjee2020,Meenakshi2023} and is further corroborated by the polarization map of J1712$-$2435 (refer to Fig.~\ref{fig:Inner_Jet_pol}, \ref{fig:South_pol}). This also aligns with our model, where we injected the jet using a pure toroidal B-field configuration. However, as the wind speed continues to increase further out, the second bending becomes intense enough to eventually lead to complete jet decollimation. After 75 Myr of evolution, the acquired structure nearly resembles the observed southern arm of J1712$-$2435, with decollimation occurring around 275 kpc, an initial bending at roughly 100 kpc (note the changing extents since 29 Myr, highlighting the structure’s dynamically evolving nature), and a second bending around 200 kpc, as illustrated in the 3D volume-rendered plot in Fig.~\ref{fig:south_arm}. The continuously expanding bifurcated `lobe-cocoon' structure becomes quite evident after 75 Myr of evolution.

\subsubsection{Development of the Northern Jet Structure}
In simulating the northern jet arm of J1712$-$2435 with a slightly higher wind speed \chg{(1.5 times the speed used for southern jet arm)}, the generated backflow material from the jet head begins to deflect significantly, forming a cocoon structure with an asymmetric width --- extending more broadly in the direction of the wind flow with respect to the jet beam. This suggests from the early stages that the deflected backflows are gradually forming a laterally expanded lobe structure rather than the elongated cocoon, as seen for the southern arm. During this time, the jet maintains a straight-line trajectory for up to 22 Myr, traveling nearly 140 kpc. Note that the jet beam exhibits a shorter spatial and temporal extent in maintaining linear flow (until the bending takes over) compared to the southern arm, clearly due to the influence of a higher ambient wind speed.

\begin{figure}
\centerline{\includegraphics[width=\columnwidth]{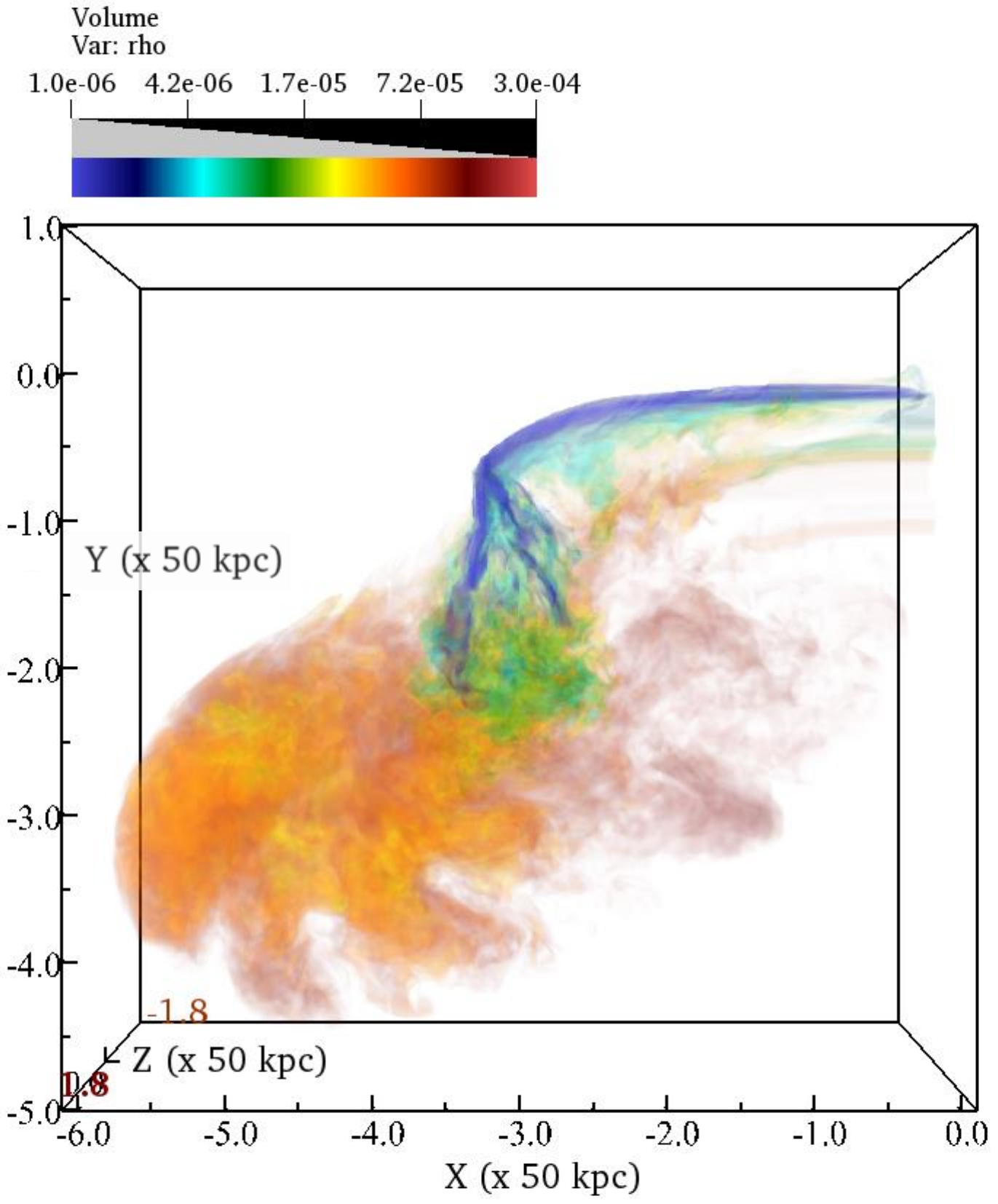}}
    \caption{Simulated 3D volume-rendered density map, obtained after 75 Myr of evolution (dynamical age) for a jet exposed to an environmental wind flow of $\omega = 0.015c/250 \, \text{kpc}$, \chg{resembling} the northern jet of J1712$-$2435. A slight increase in wind speed has produced several distinct features, including sharp jet bending and decollimation into a broad, cone-shaped structure. As a result of this pronounced bending, the jet now expands more laterally than along its original flow direction. \chg{The density scale in the colorbar is presented}  with respect to 0.001 amu/cc.}
\label{fig:north_arm}
\end{figure}

Following the initial bending, the jet attempts to stabilize against further distortion, resulting in numerous small-scale twists, turns, and partial detachment of the jet beam near the jet head. This process continues for an additional 40 Myr, during which the lobe structure steadily develops. By a dynamical age of nearly 65 Myr, the jet undergoes a significant bend, almost perpendicular to its original propagation direction, infusing the lobe structure with additional jet material and forming well-defined boundaries at the jet-ambient interface. At this stage, the jet-head decollimates into multiple finger-like structures. With further evolution, the bending and decollimation intensify, ultimately producing the structure depicted in Fig.~\ref{fig:north_arm}. The lateral deflection of the jet beam, its decollimation into a broad cone-shaped structure, and the partially well-defined boundaries on one side, with tendril-like extensions on the other, are particularly evident in this morphology.

\begin{figure}
\centerline{\includegraphics[width=\columnwidth]{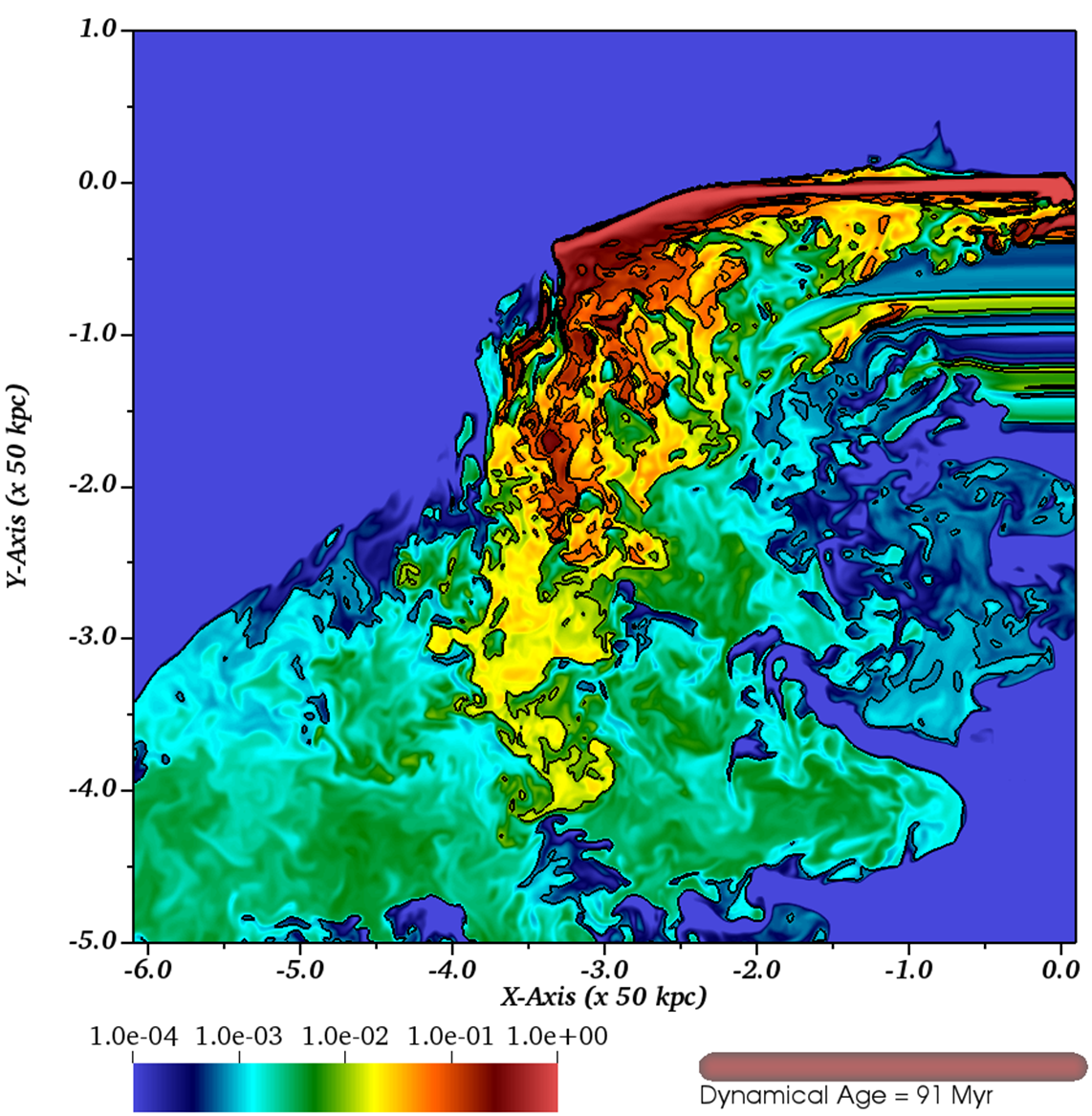}}
    \caption{2D slice ($x-y$ plane, $z = 0$) of the resulting structure (as visualized through tracer values) from the simulation of the northern jet arm of J1712$-$2435 at a dynamical age of 91 Myr. The colormap illustrates the distribution of tracers, highlighting the jet material and its fraction, with tracer contours overlaid to emphasize the dominant jet-flow regions. The frustum-like extended lobe morphology is particularly evident here, closely resembling the structure of the northern arm of J1712$-$2435, followed by broad tendril-like extensions.}
\label{fig:north_arm_sliced}
\end{figure}

Due to the \chg{similarity in} morphological appearance in the simulated structure modeling the northern jet topology (i.e., Fig.~\ref{fig:north_arm}), we extended this simulation to observe its evolution over a longer timescale. The structure obtained at a dynamical age of 91 Myr is showcased in Fig.~\ref{fig:north_arm_sliced}. This 2D-sliced image highlights again the sudden termination of the jet flow due to the increased external wind speed, leading to matter being stripped from the jet head, diffusing laterally, and filling the evolving lobe. The morphology resembles a frustum-shaped cone with tendril-like extensions growing from the lobe. This structure closely resembles the northern arm of J1712$-$2435, further illustrating that our incorporated physical models are highly relevant for explaining the distinct tailed morphology of J1712$-$2435.

\subsubsection{A Combined Overview and Caveats} \label{Sec:A Combined Overview and Caveats}
\begin{figure*}
\centerline{\includegraphics[width=2.\columnwidth]{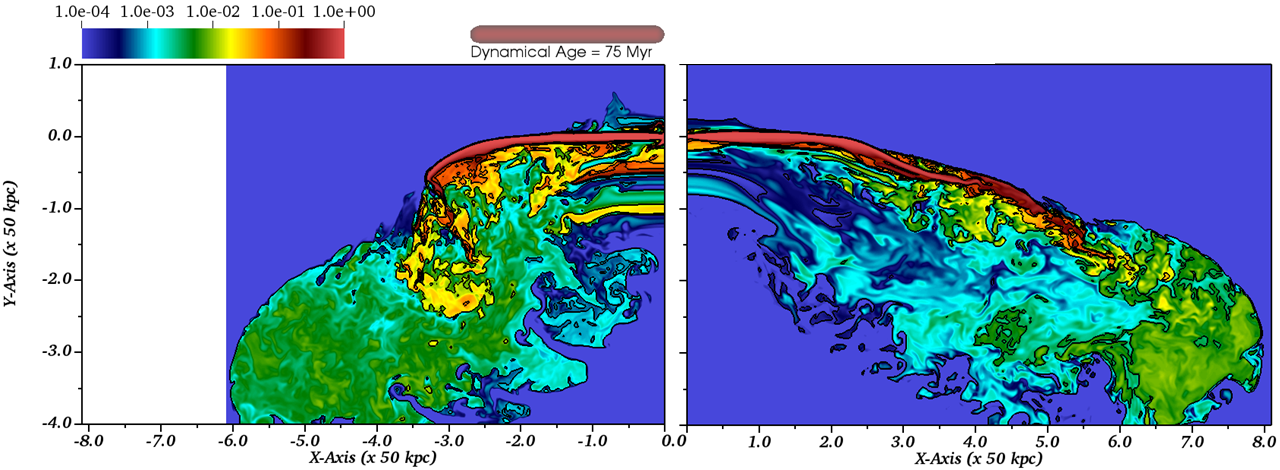}}
    \caption{2D slice ($x-y$ plane, $z = 0$) of the simulated 3D tracer distribution is presented in a color scale at a dynamical age of 75 Myr, mimicking the jet-lobe structure in the `northern' and `southern' arms of J1712$-$2435. \chg{This visualization underscores the competing roles of the sheltering effect of the host galaxy gas and the environmental wind in reproducing the observed source characteristics.}  To enhance clarity, the simulation intended for the southern arm has been mirrored along the Y-axis. Both arms are shown over an $x$-direction extent of 400 kpc, illustrating the distinct evolutionary behaviors resulting from a \chg{marginal} variation in the environmental wind speed \chg{(velocity \chg{ratio} of $\sim 1.5$ times)}. Some caveats and potential areas for further improvement of the models can be found in Section~\ref{Sec:A Combined Overview and Caveats}.}
\label{fig:SN_combined}
\end{figure*}

Fig.~\ref{fig:SN_combined} presents a simplified depiction of the simulated jet-lobe structures for the southern and northern arms of J1712$-$2435 at 75 Myr. To facilitate comparison, one arm has been mirrored, while the other is plotted over the same spatial extent. The figure highlights the notably different behaviors that emerge due to a minimal change in the speed of the ambient medium (velocity \chg{ratio} of $\sim 1.5$ times). It is worth noting that the initial propagation of the jet (in spatial terms) remains stable and maintains a straight-line trajectory. This stability is due to the lower linear speed of the environmental wind flow, which effectively mimics the influence of the host galaxy's interstellar medium in shielding the jet during its early spatial phase. The arms of J1712$-$2435 exhibit their first bending at a distance of 98 kpc. In our simulated models, the bending begins gradually around 100 kpc and then becomes more pronounced, further supporting the model's relevance and the assumptions made for the source. Such extents in the jet-to-tail transition are not uncommon \citep{ODea2023}, suggesting that once the jet exits the nurturing influence of the host galaxy \chg{(i.e., the partial shielding provided by galactic gas from the external hazardous environment)}, \chg{the intergalactic wind is sufficiently strong to shape the tail morphology in accordance with its flow speed (Fig.~\ref{fig:SN_combined}).}

It is important to emphasize that, despite the formation of tendril-like extensions from the simulated northern jet-arm's lobe structure --- which may signify the early stages of filamentary features observed in the northern arm of J1712$-$2435 --- these extensions struggle to reach the linear scales observed in J1712$-$2435. This suggests that reproducing the full extent of filamentary structures seen in this source likely requires more complex modeling, such as the incorporation of magnetic fields in the intergalactic medium and enhanced shear magnitude between the radio lobe and surrounding medium \citep[][]{Rudnick2022}. 
\chgb{A potential hint for such a scenario can be understood by the small, yet necessary, difference in environmental wind speed required to reproduce the observed morphology of both jet arms. Although one would typically expect the jets of J1712$-$2435 to interact with a nearly symmetric ambient wind, our modelling indicates that a mismatch in wind magnitude improves the reproduction of structural asymmetries. This suggests that additional factors—beyond simple hydrodynamical effects—may be influencing jet propagation. One such factor could be the presence of an asymmetric environmental magnetic field, which, despite being typically considered dynamically subdominant \citep{Stuardi2020}, may exert a non-negligible influence on jet evolution under certain conditions. Asymmetric field distributions are a natural outcome in merging or dynamically disturbed large-scale environments, and recent numerical-observational synergy studies \citep[e.g.,][]{Chibueze2021} have begun to explore their potential impact on jet-environment interactions. This scenario also provides a plausible physical context for the formation of the extended filamentary structures observed in J1712$-$2435, which remain difficult to explain through pure hydrodynamic effects alone, as indicated by our current simulations.}
Understanding the mechanisms behind the formation of elongated filament-like structures emerging from the radio lobe boundaries and extending outward is a frontier topic of research, especially with several recent observations of similar features in other sources \citep[e.g.,][]{Ramatsoku2020,Condon2021,Velovic2023,Koribalski2024}. \chgb{ While our current modeling does not explicitly incorporate a detailed treatment of ambient magnetic field line distributions and their spatial variation, we recognize that such components — particularly in a resistive-MHD framework that allows for magnetic reconnection \citep{DelSarto2003,Mignone2019}— may play a key role in the development of these structures. A more dedicated numerical investigation incorporating such effects would be essential to more fully capture the physical processes shaping the observed filamentary features.}

We also note that our results are marginally impacted by boundary effects at the jet injection side (right boundary along the $x-$axis). To simulate a subset of the rotating ambient medium, we applied outflow boundaries on all sides, which incidentally allowed some inward flow \chg{of environmental matter} near the right $x-$boundary. This inward flow has a limited effect on our main findings, aside from giving the appearance that the jet beam is accompanied by cocoon material in the vicinity of this boundary. While this effect is distinguishable, it can be ignored; due to the high computational cost of fully mitigating it, we chose to retain the current setup.

\chg{Among the parameter variations explored in this study while modeling J1712–2435, it remains a testable hypothesis whether a combination of higher ambient gas density and lower wind speed could reproduce the observed morphology. However, pursuing this scenario would require a new suite of computationally intensive simulations, which we have chosen not to undertake at this stage. \chgb{This decision reflects the current lack of compelling observational support—for instance, the uncertain evidence for the host galaxy’s location within a rich cluster environment} (see Section~\ref{Sec:Host Galaxy and Environment}). Additionally, the morphological traits typically associated with WATs, such as the gradual bending of the jet arms, are in contrast with the sharp and complex deflections observed in J1712$–$2435, requiring the assumption of a higher wind speed.}

\chgb{However, as the wind speed required in our model lies toward the higher end of values typically reported in the literature \citep{Musoke2020,ODea2023}, we also discussed a set of alternative scenarios that may account for the complex morphology observed in J1712$–$2435 (Section~\ref{Sec:Alternate Scenario}). These scenarios—including the one presented in this work—offer physically plausible interpretations but currently lack direct observational validation. We therefore underscore the importance of future observational campaigns focused on characterizing the large-scale environment around J1712$-$2435, which will be crucial for disentangling the dominant physical mechanisms shaping its structure.}

\subsubsection{\chg{Alternate Scenario}}\label{Sec:Alternate Scenario}
\chg{As an alternative to the environmental wind-driven bending model proposed in this work, one may consider a scenario where the morphology of J1712--2435 results from the combined influence of buoyancy forces, arising due to the pressure gradient within the host galaxy group, and magnetohydrodynamic instabilities in the jet beam, such as kink and pinch modes induced by magnetic fields \citep{Cowie1975}. The interaction of these effects could potentially reproduce several key features of the source: for instance, the multiple bends observed in the southern jet \citep[e.g.,][]{Mignone2013,Mukherjee2020,Acharya2021}, the subsequent decollimation and formation of lobes, and the one-sidedness of the jet-induced lobes as a consequence of buoyant motion through a stratified ambient medium \citep[e.g.,][]{Burns1982,Sakelliou1996}. While this combined mechanism may offer a plausible physical interpretation, it also raises a number of issues. It remains unclear why two jets launched under the same injection conditions would evolve so differently in terms of structure and stability. Additionally, such a model must account for the origin of filamentary features, the bifurcation observed in the southern lobe, and the formation of a northern lobe in the absence of significant jet decollimation. \chgb{Addressing these questions would require a large and computationally intensive parameter space exploration, which may be mitigated by observational constraints on the relevant physical parameters.}}
\chgb{We therefore reiterate the importance of future observational campaigns focused on characterizing the large-scale environment around J1712$-$2435.}

\subsection{Radial Expansion of Jets with Propagation Distance}
\begin{figure*}
\centerline{\includegraphics[width=1.8\columnwidth]{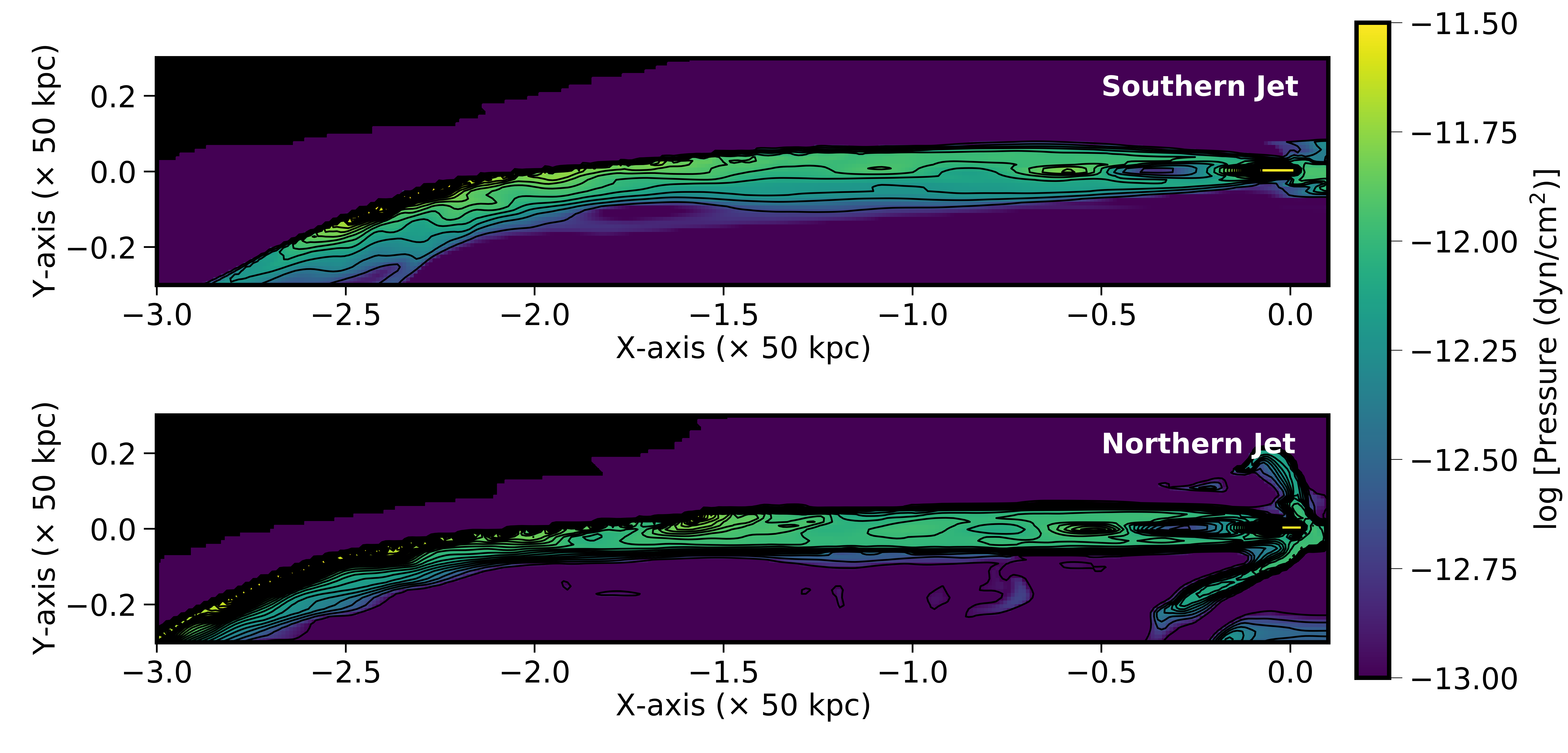}}
    \caption{Sliced distribution ($x-y$, $z = 0$) of thermal pressure, illustrating the widening of the jets’ cylindrical flow since their injection, shaped by their propagation through a marginally under-pressured ambient medium. The thermal pressure colormap is overlaid with black pressure contours to visually emphasize the radial expansion of the lateral jet extent, while also highlighting regions of pressure compression and rarefaction. The plot spans up to 150 kpc, capturing the jets' behavior near their first bend at approximately 100 kpc.}
\label{fig:Pressure_slice}
\end{figure*}

We investigated the behavior of the jet beams as they propagate through the varying pressure distribution of the ambient medium. The jets were launched into the simulation domain with an initial pressure nearly equivalent to that of the surrounding medium. \chg{However, as the jets propagate outward, the average jet-to-ambient pressure ratio slowly increases from approximately 1.1 to 1.9, driven by the gradual decline in ambient pressure (as set by the initial conditions) and a modest rise in jet pressure due to resistance to propagation.} This over-pressured nature of the jets drives their radial expansion, leading to noticeable lateral widening. This explanation is illustrated in Fig.~\ref{fig:Pressure_slice}, which presents a zoomed-in view of the sliced pressure distribution, highlighting the spatial evolution of the jet beams for both the southern and northern jets at 75 Myr. The lateral expansion is evident and can be clearly discerned through the overlaid pressure contours.

One notable feature is the formation of pressure compression zones along the jet beam, predominantly in the base region where the jet remains straight, i.e., within the initial spatial extent from the launching zone. These compressed regions are expected to exhibit enhanced emission, which could explain the observed excess emissions in the jet base region of J1712$-$2435. Another notable feature is the enhancement of pressure values in the jet beam as it bends under the influence of an external wind around 100 kpc. In the northern jet, this pressure enhancement is more widespread due to the stronger wind, resulting in a pronounced bend that could explain the lit-up region at the jet-beam termination in the northern arm of J1712$-$2435. In contrast, the southern jet is only affected mildly due to weaker external wind speed, leading to the pressure enhancement confined to a smaller zone near the bend, which may account for the absence of a lit-up region in the bends of southern arm. Additionally, as the external wind typically flows from the top of the simulation domain to the bottom, a pressure gradient forms in the jet-beams, with the highest pressure at the top-side of the jet and a gradual decrease downward (see, Fig.~\ref{fig:Pressure_slice}).

\begin{figure}
\centerline{\includegraphics[width=\columnwidth]{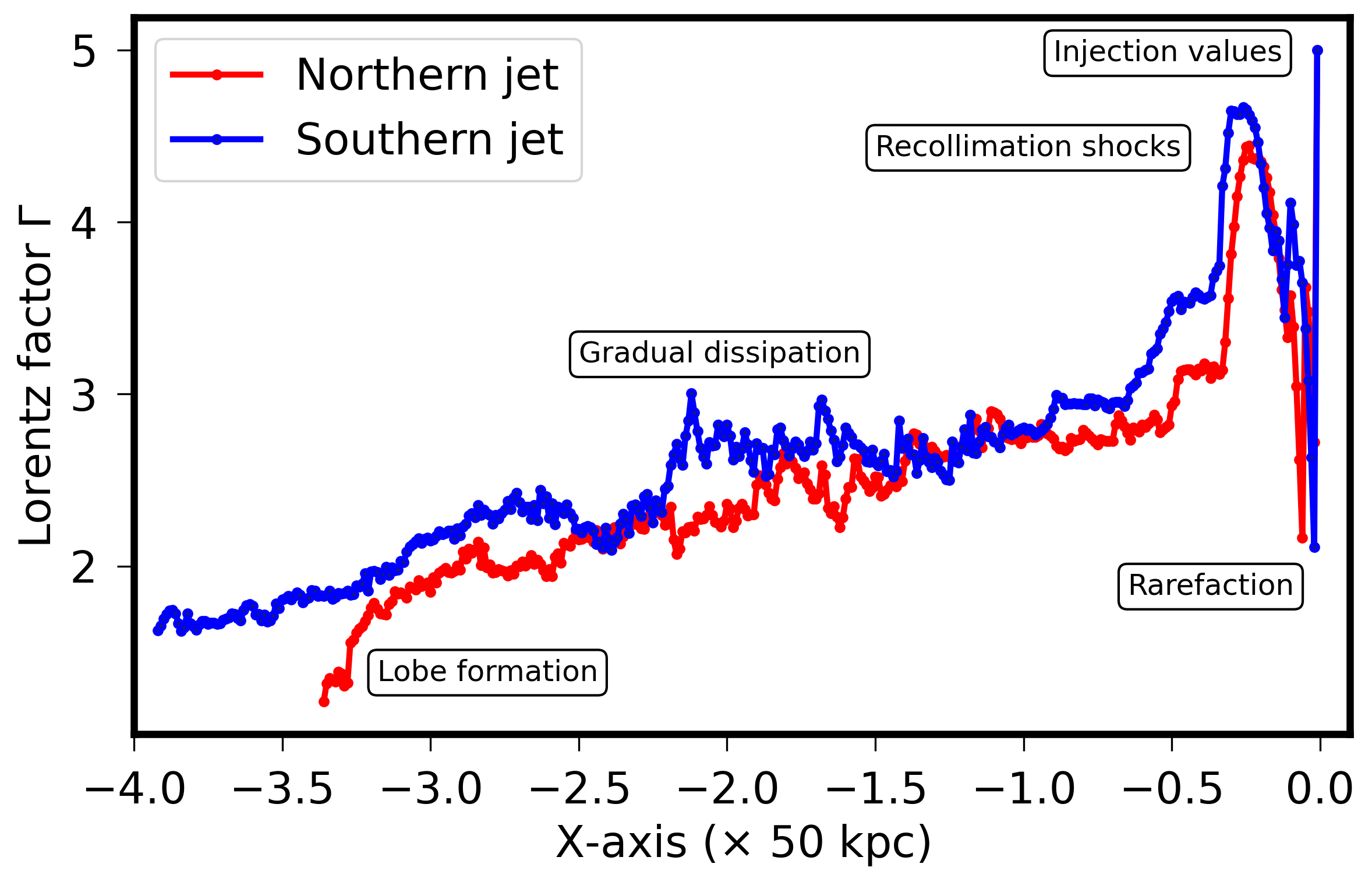}}
    \caption{Variation of median \chg{(bulk)} Lorentz factor ($\Gamma$) values along the jet spine up to the initial lobe formation region. The injection condition for $\Gamma$ is set to 5 in our simulations, followed by a recollimation shock zone. Subsequently, a gradual decrease in propagation speed is observed for both arms. While the southern arm maintains its collimation and propagation, the northern arm decollimates into lobes after a sharp bend, dissipates energy, and transitions into the subrelativistic regime.}
\label{fig:LorG}
\end{figure}

As the jet flows in the plane of the sky, the Doppler factors derived from our simulations are close to unity, indicating that Doppler effects have minimal impact on the observed appearance of the jet. In this context, an analysis of the jets' Lorentz factor ($\Gamma$) distributions reveals that these jets actively propagate up to their first bends ($\sim 100$ kpc) with median Lorentz factor values of 2.5 and above (compared to an injection value of 5). However, as soon as the lobe formation starts, followed by bending, the material dissipates its mechanical energy over a broader angle, entering into subrelativistic evolution by drastically decreasing the $\Gamma$ values. These dissipation phases are highlighted in Fig.~\ref{fig:LorG}, where, immediately after injection, the detection of rarefaction followed by compression zones has also been observed (indications of recollimation shocks; see, Fig.~\ref{fig:Pressure_slice}). Such a shock likely explains the enhanced emission zones generated near the core of J1712$-$2435 (Fig.~\ref{fig:Full_IPol}). This behavior further supports the conclusion that the host galaxy's sheltering effect plays a significant role in the collimation and initial propagation of tailed radio galaxies. In this context, a recent study by \citet{Rossi2024} performed to understand the initial formation phases of FR-I type sources has highlighted a similar $\Gamma$ distribution in their simulated maps for cases with low magnetizations, relevant to our injection values.

\begin{figure}
\centerline{\includegraphics[width=\columnwidth]{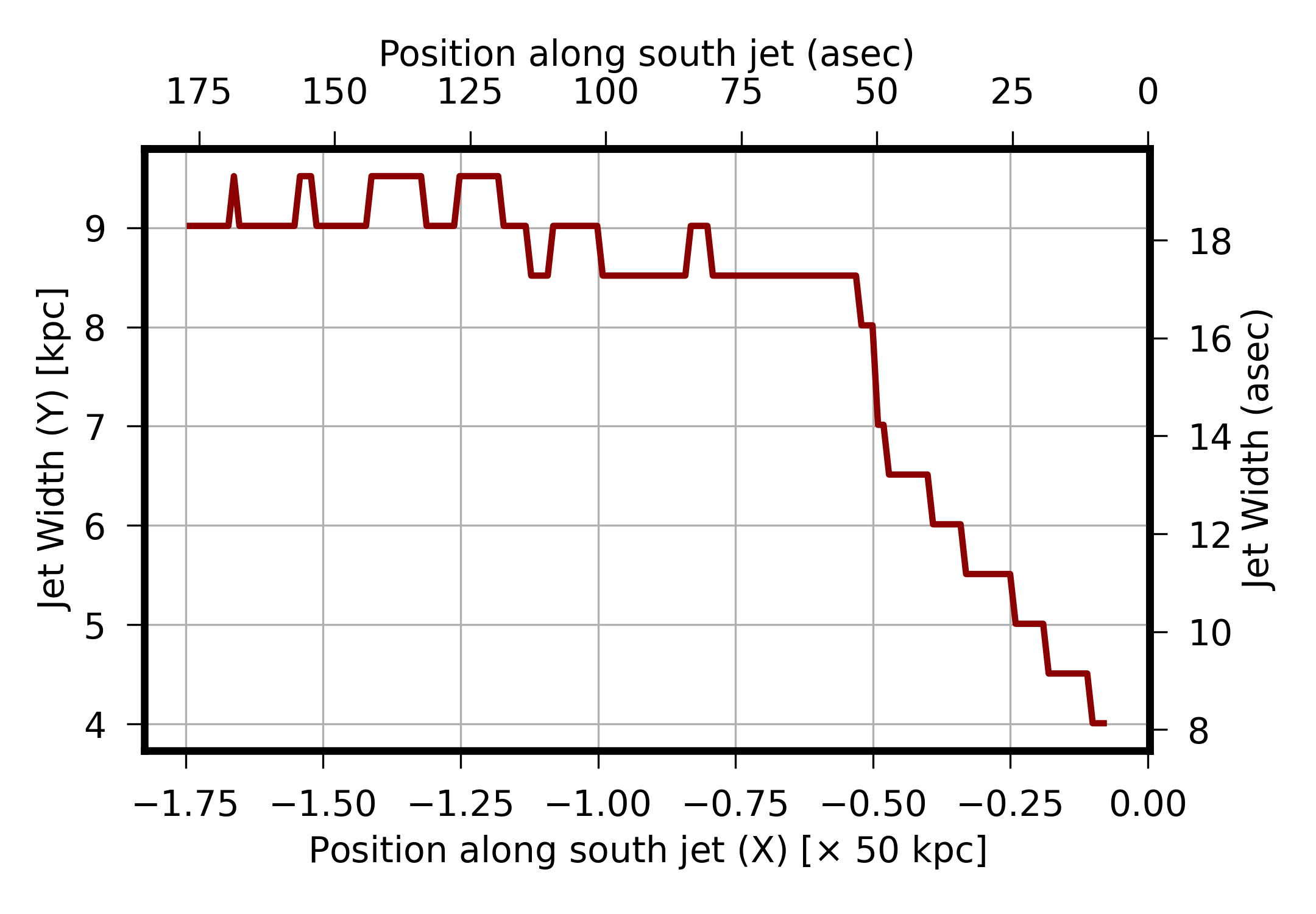}}
    \caption{\chg{Illustration of the lateral expansion of the simulated southern jet as a function of distance along the propagation axis. The analysis is based on the simulation snapshot at 75 Myr, aiming to quantify the transverse jet growth visible in Fig.~\ref{fig:Pressure_slice}. The figure reveals a phase of rapid initial lateral expansion, which transitions to a more gradual widening closely aligning with the location of a recollimation shock, suggesting its role in regulating the jet’s cross-sectional growth and decollimation.}}
\label{fig:Jet_width_length}
\end{figure}

\chg{We further investigated the evolution of jet width with increasing distance from the injection point to quantify the influence of the recollimation zones on the jet's lateral behavior. We present the behavior of the simulated southern jet in Fig.~\ref{fig:Jet_width_length}, as it represents a clean jet system and allows for a direct comparison with observational data shown in Fig.~\ref{fig:JetExpansion}. A rapid lateral expansion of the jet is observed within the initial $\sim 25$ kpc (corresponding to $\sim 50$ arcsecond; $z = 0.024330$), beyond which the expansion transitions to a more gradual rate. When compared with Fig.~\ref{fig:LorG}, this transition appears to occur shortly after the onset of the recollimation shock, suggesting that the jet undergoes collimation which curtails further rapid expansion and decollimation. Fig.~\ref{fig:JetExpansion}, derived from observational data, displays a resembling trend in the jet's width evolution.}

\section{Summary\label{Summary}}
We present observations of the \chg{radio}\chgb{-}\chg{loud} GRG J1712$-$2435 with well
collimated jets with minor bends before becoming decollimated and
forming extended plumes.
The \chg{projected length} of the source is approximately 1 Mpc with bends and 
decollimation starting at around 100 kpc from the nucleus.
Due to its proximity to the Galactic
Center (l$\sim$359.6$^\circ$, b$\sim$8.5$^\circ$), the detailed
environment is not well understood but there are known to be massive
clusters nearby.
The WAT structure of the source shows an intergalactic medium capable of
decollimating and deflecting the jets.

The brightness ratio of the inner jets suggest that their plane is nearly in that of the sky and an estimate of the jet velocity cannot be made.  The 1400 MHz power is 3.9$\times 10^{24}$ W Hz$^{-1}$, slightly below the FRI/FRII divide \citep{Owen1994} at \chg{$\sim 10^{25}$ W Hz$^{-1}$}.  The total power radiated between 10 MHz and 100 GHz is estimated at 5.6$\times 10^{41}$ erg sec$^{-1}$. Although the source is nearly in the plane of the sky, the large ratio of \chg{projected jet lengths} of $\sim$2.8 suggests that the external environment is shaping this along with the overall structure, assuming the jets to be intrinsically symmetric.

We have simulated the overall behavior of this wide\chgb{-}angle tail source using magneto-hydrodynamical modeling with a variable intergalactic wind speed.  In order to mock the sheltering effect on the host galaxy ISM, a rotating external medium is used with a separate rotational model for independent simulations of the northern and southern jets.  These simulations reproduce the gross morphology of J1712$-$2435 and \chgb{suggest} the role of the host galaxy's sheltering effect, along with the jet's magnetic field configuration, \chgb{to shape} the collimation and spatial extent of tailed radio galaxies.

\chgb{This study also outlines alternative scenarios that merit further investigation, highlighting the need for coordinated observational and numerical efforts to unravel the origin of such a distinctive source, shaped by the complex conditions of its large-scale environment.}

\section{Data \chg{Availability}\label{Products}}
The raw data is available from the SARAO archive
(\url{https://archive.sarao.ac.za/}) under project code SSV-20200423-FC-01. 
Image products can be obtained from doi (\url{https://doi.org/10.484792/drvw-vz92}).

The broadband images cubes have names of the form
``J1712-2435\_$<$S$>$Pol.fits.gz'' where $<$S$>$ is I, Q, U or V and have the form
described in \cite{MFImage}.
The second plane of the Stokes I image \chg{cube} is the spectral index and the
second plane of the other Stokes types is blanked.

There is a second form of the Stokes I image,
J1712-2435\_I\_FitSpec.fits.gz, in which the first two planes are the
same as in the Stokes I cube followed by 3) the error estimate for
total intensity (where fitted), 4) the error estimate of spectral
index and 5) the reduced $\chi^2$ of the fit. 
The reference frequency of all broadband images is 1333.13 MHz and
Stokes I broadband images have been primary beam corrected.
Note: the broadband Q and U images have not been corrected for Faraday
rotation and the EVPA at $\lambda=0$ and polarized intensity from the
Faraday analysis cube should be used instead.
None of the subband planes in the image cubes are primary beam
corrected.
\chg{Channelization of the images is detailed in the Appendix.}

The result of the Faraday analysis are in a cube,
J1712-2435\_RM.fits.gz, with planes 1) Faraday depth (rotation
measure) of the peak unwrapped polarized intensity (rad. m$^{-2}$), 
2) the EVPA at $\lambda=0$ (rad.), 3) unwrapped polarized intensity
(Jy), 4) the reduced $\chi^2$ of the fit. 
The first two planes of the J1712-2435\_RM.fits.gz image are blanked
where the polarized intensity (3rd plane) is below 35 $\mu$Jy
beam$^{-1}$.

\appendix
\section{Subband Frequencies}
The Stokes I and V image cubes have the channelization described in Table
\ref{tab:subband} while planes 3-70 in the Q and U cubes have center
frequencies (MHz) of 
890.482, 899.260, 908.037, 916.814, 925.592,
934.787, 943.564*, 952.342*, 961.119, 969.896,
979.092, 987.869, 997.064, 1007.096, 1017.127,
1026.740, 1036.354, 1046.385, 1056.416, 1066.447,
1076.061, 1085.674, 1096.123, 1106.990, 1117.857,
1128.725, 1139.592, 1150.459, 1161.326*, 1172.193*,
1183.061*, 1194.346*, 1206.049*, 1218.170*, 1230.291*,
1241.994*, 1253.697*, 1265.400*, 1277.104*, 1289.225*,
1301.764, 1314.303, 1327.260, 1340.217, 1352.756,
1365.713, 1378.670, 1392.045, 1405.838, 1419.631,
1433.424, 1447.217, 1461.010, 1474.803, 1489.014,
1503.225, 1517.854, 1532.900*, 1547.529*, 1562.158*,
1577.205*, 1592.670*, 1608.553, 1624.436, 1640.318,
1656.201, 1672.084, 1680.861*.
Subbands which are totally flagged due to RFI are marked with ``*".
\begin{table}[h]
    \centering
    \begin{tabular}{c|c|c}
    Subband & Frequency & Comment \\
             & MHz      & \\
             \hline
       0 &  1333.13& Broadband\\
       2 &  908.0 & \\
       3 &  952.3 & \\
       4 &  996.6 & \\
       5 & 1043.4 & \\
       6 & 1092.8 & \\
       7 & 1144.6 & \\
       8 & 1198.9 & Blanked\\
       9 & 1255.8 & Blanked\\
       10 & 1317.2 & \\
       11 & 1381.2 & \\
       12 & 1448.1 & \\
       13 & 1519.9 & \\
       14 & 1593.9 & \\
       15 & 1656.2 & \\
       \hline
    \end{tabular}
    \caption{MeerKAT sub-band central frequencies.}
    \label{tab:subband}
\end{table}

\section*{Acknowledgments}
We would like to thank Alan Bridle for an informative discussion. GG is a postdoctoral fellow under the sponsorship of the South African Radio Astronomy Observatory (SARAO). The financial assistance of the SARAO towards this research is hereby acknowledged.
The MeerKAT telescope is operated by the South African Radio Astronomy
Observatory, which is a facility of the National Research Foundation,
an agency of the Department of Science and Innovation. 
The National Radio Astronomy Observatory and Green Bank Observatory are facilities of the National
Science Foundation, operated under a cooperative agreement by
Associated Universities, Inc.
We acknowledge the use of the ilifu cloud computing facility - \url{https://www.ilifu.ac.za/}, a partnership between the University of Cape Town, the University of the Western Cape, Stellenbosch University, Sol Plaatje University, the Cape Peninsula University of Technology and the South African Radio Astronomy Observatory. The ilifu facility is supported by contributions from the Inter-University Institute for Data Intensive Astronomy (IDIA - a partnership between the University of Cape Town, the University of Pretoria and the University of the Western Cape), the Computational Biology division at UCT and the Data Intensive Research Initiative of South Africa (DIRISA).
This research has made use of the SIMBAD database,
operated at CDS, Strasbourg, France.
This research has made use of the NASA/IPAC Extragalactic Database,
which is funded by the National Aeronautics and Space Administration
and operated by the California Institute of Technology.  

\vspace{5mm}
\facilities{MeerKAT}

\software{Obit \cite{OBIT}}
\bibliography{J1712-2435}{}
\bibliographystyle{aasjournal}


\end{document}